\documentclass[11pt]{article}
\usepackage[dvips]{graphicx}
\begin{document}
\title{\vspace*{-3em} {\normalsize
This document can be cited as: Draine, B.T., and Flatau, P.J. 2000, \\
``User Guide for the Discrete Dipole Approximation Code DDSCAT 
(Version 5a10)'', \\
\vspace*{-0.6em} http://arxiv.org/abs/astro-ph/0008151v4} \\ \vspace*{2em}
	{\bf User Guide for the Discrete Dipole} \\
	{\bf Approximation Code {{\bf DDSCAT}}}\\
	(Version 5a10)}
                                
\author{Bruce T. Draine \\
	Princeton University Observatory \\
	Princeton NJ 08544-1001 \\
	({\tt draine@astro.princeton.edu})\\
	\ \\
	and \\
	\ \\
	Piotr J. Flatau \\
        University of California, San Diego \\
	Scripps Institution of Oceanography \\
	La Jolla CA 92093-0221 \\
	({\tt pflatau@ucsd.edu})
	}
\date{last revised: 2002 May 31}
\maketitle
\abstract{
	{{\bf DDSCAT.5a}} is a freely available 
	software package which applies the
	``discrete dipole approximation'' (DDA) to calculate scattering
	and absorption of electromagnetic waves by targets with arbitrary
	geometries and complex refractive index.  The DDA approximates
	the target by an array of polarizable points.
	{{\bf DDSCAT.5a}} requires that these polarizable points be located
	on a cubic lattice.  {{\bf DDSCAT.5a10}} allows
	accurate calculations of electromagnetic scattering from targets
	with ``size parameters'' $2\pi a/\lambda < 15$ provided the
	refractive index $m$ is not large compared to unity ($|m-1| < 1$).

	The {{\bf DDSCAT}} package is written in Fortran 
	and is highly portable.
	The program supports calculations for a variety of target geometries
	(e.g., ellipsoids, regular tetrahedra, rectangular solids, 
	finite cylinders, hexagonal prisms, etc.).
	Target materials may be both inhomogeneous and anisotropic.
	It is straightforward for the user to ``import'' arbitrary
	target geometries into the code, and relatively straightforward
	to add new target generation capability to the package.
	{{\bf DDSCAT}} automatically calculates total cross sections
	for absorption and scattering and selected elements of the
	Mueller scattering intensity 
	matrix for
	specified orientation of the target relative to the incident wave,
	and for specified scattering directions.

	This User Guide explains how to use {{\bf DDSCAT.5a10}} to carry
	out electromagnetic scattering calculations.  
	CPU and memory requirements are described.
	}
\newpage
\tableofcontents
\newpage
\section{Introduction\label{intro}}
{{\bf DDSCAT.5a}}\ is a Fortran software package to calculate scattering and
absorption of electromagnetic waves by targets with arbitrary geometries
using the ``discrete dipole approximation'' (DDA).
In this approximation the target is replaced by an array of point dipoles
(or, more precisely, polarizable points); the electromagnetic scattering
problem for an incident periodic wave interacting with this array of
point dipoles is then solved essentially exactly.
The DDA (sometimes referred to as the ``coupled dipole approximation'') 
was apparently first proposed by Purcell \& Pennypacker (1973).
DDA theory was reviewed and developed further by Draine (1988), 
Draine \& Goodman (1993), and recently reviewed by Draine \& Flatau (1994)
and Draine (2000).

{{\bf DDSCAT.5a}}\ is a Fortran implementation of the DDA developed by the authors.
The previous version, {\tt DDSCAT.5a9}, was released 1998 December 15.
The current version, {\tt DDSCAT.5a10}, released 2000 June 15, adds a
new ``multisphere'' target option.
{{\bf DDSCAT.5a}}\ is 
intended to be a versatile tool, suitable for a wide variety
of applications ranging from interstellar dust to atmospheric aerosols.
As provided, {{\bf DDSCAT.5a10}}\ should be usable for many 
applications without
modification, but the program is written in a modular form, so that
modifications, if required, should be fairly straightforward.

The authors make this code openly available to others, in the hope that it
will prove a useful tool.  We ask only that:
\begin{itemize}
\item If you publish results obtained using {{\bf DDSCAT}}, please consider 
	acknowledging the source of the code.

\item If you discover any errors in the code or documentation, 
	please promptly communicate them to the authors.

\item You comply with the ``copyleft" agreement (more formally, the 
	GNU General Public License) of the Free Software Foundation: you may 
	copy, distribute, and/or modify the software identified as coming 
	under this agreement. 
	If you distribute copies of this software, you must give the 
	recipients all the rights which you have. 
	See the file {\tt doc/copyleft} distributed with the DDSCAT software.
\end{itemize}
We also strongly encourage you to send email to the authors identifying  
yourself as a user of DDSCAT;  this will enable the authors to notify you of
any bugs, corrections, or improvements in DDSCAT.

The current version, {{\bf DDSCAT.5a10}}\ , 
uses the DDA formulae from Draine (1988).
The code incorporates Fast Fourier Transform (FFT) methods
(Goodman, Draine, \& Flatau 1991).
The ``lattice dispersion relation'' (LDR) prescription (Draine \& Goodman 1993)
is used for determining dipole polarizabilities.

This User Guide assumes that you have already obtained the Fortran
source code for {{\bf DDSCAT.5a10}}\ either via the World Wide Web
({\tt http://www.astro.princeton.edu/}$\sim${\tt draine}) or via
anonymous {\tt ftp} following the instructions in the {\tt README}
file.\footnote{To 
	obtain the {\tt README} file: (1) anonymous {\tt
	ftp} to {\tt astro.princeton.edu} , (2) {\tt cd
	draine/scat/DDA/ver5a}, and (3) {\tt get README}.
	}
We refer you to the list of references at the end of this document for
discussions of the theory and accuracy of the DDA [first see the
recent reviews by Draine and Flatau (1994) and Draine (2000)].  In
\S\ref{sec:whats_new} we describe the principal changes between {{\bf
DDSCAT.5a10}}\ and the previous releases.\footnote{
	The first ``official release'' was {\bf DDSCAT.4b}, although
	{\bf DDSCAT.4c} -- while never announced -- was made available to a
	number of interested users.
	{\bf DDSCAT.5a8} was released in 1997 April.
	{\bf DDSCAT.5a9} was released in 1998 December.
	{\bf DDSCAT.5a10} was released 2000 June 15.
	}
The succeeding sections contain instructions for:
\begin{itemize}
\item compiling and linking the code;
\item running a sample calculation;
\item understanding the output from the sample calculation;
\item modifying the parameter file to do your desired calculations;
\item specifying target orientation;
\item changing the {\tt DIMENSION}ing of the source code to accommodate 
your desired calculations.
\end{itemize}
The instructions for compiling, linking, and running will be appropriate for a
UNIX system; slight changes will be necessary for non-UNIX sites, but they are
quite minor and should present no difficulty.

Finally, this User Guide can be obtained by any of the following means:
\begin{itemize}
\item {\tt http://xxx.lanl.gov/abs/astro-ph/0008151} -- you will be offered
the options of downloading 
	\begin{itemize}
	\item 
	Latex source
	\item 
	Postscript
	\item 
	Other formats -- click on this to obtain the UserGuide as a PDF file.
	\end{itemize}
\item anonymous {\tt ftp} from
{\tt astro.princeton.edu}, subdirectory {\tt draine/scat/ddscat/ver5a10}.
	\begin{itemize}
	\item 
	For postscript, {\tt get} either the 
	uncompressed postscript file 
	{\tt UserGuide.ps} or the gzipped file {\tt UserGuide.ps.gz} 
	(remember to set the {\tt binary} option in {\tt ftp} before using 
	the {\tt get} command).
	\item 
	Alternatively, you can {\tt get} the LaTeX source 
	{\tt UserGuide.tex.gz} and
	the figures {\tt f1.eps.gz}, {\tt f2.eps.gz}, {\tt f3.eps.gz}, 
	{\tt f4.eps.gz}, and {\tt f5.eps.gz}.
	\end{itemize}
\end{itemize}

\section{Applicability of the DDA\label{sec:applicability}}
The principal advantage of the DDA is that it is completely flexible 
regarding the geometry of the target, being limited only by the need to 
use an interdipole separation $d$ small compared to 
(1) any structural lengths in the target, and
(2) the wavelength $\lambda$.
Numerical studies (Draine \& Goodman 1993; Draine \& Flatau 1994; Draine 2000)
indicate that the
second criterion is adequately satisfied for calculations of
total cross sections if
\begin{equation}
|m|kd< 1~~~,
\label{eq:mkd_max}
\end{equation}
where $m$ is the complex refractive index of the target
material, and $k\equiv2\pi/\lambda$, where $\lambda$ is the wavelength
{\it in the surrounding medium} (normally taken to be vacuum, but see
\S\ref{sec:target_in_medium}).
However, if accurate calculations of the scattering phase function
(e.g., radar or lidar cross sections)
are desired,
a more conservative criterion 
\begin{equation}
|m|kd < 0.5
\end{equation}
will ensure that differential scattering cross sections
$dC_{\rm sca}/d\Omega$ are accurate to within a few percent of the
average differential scattering cross section $C_{\rm sca}/4\pi$
(see Draine 2000).

Let $V$ be the target volume.
If the target is represented by an array of $N$ dipoles, located on
a cubic lattice with lattice spacing $d$,
then 
\begin{equation}
V=Nd^3 ~~~.
\end{equation}
We characterize the size of the target by the ``effective radius''
\begin{equation}
a_{\rm eff}\equiv(3V/4\pi)^{1/3} ~~~,
\end{equation}
the radius of an equal volume sphere.
A given scattering problem is then characterized by the
dimensionless ``size parameter''
\begin{equation}
x\equiv ka_{\rm eff} = \frac{2\pi a_{\rm eff}}{\lambda} ~~~.
\end{equation}
The size parameter can be related to $N$ and $|m|kd$:
\begin{equation}
x\equiv{2\pi a_{\rm eff}\over\lambda} =
{62.04\over|m|}\left({N\over10^6}\right)^{1/3} \cdot |m|kd ~~~.
\end{equation}
Equivalently, the target size can be written
\begin{equation}
a_{\rm eff} = 9.873 {\lambda\over|m|}\left({N\over10^6}\right)^{1/3}
\cdot |m|kd~~~.
\end{equation}
Practical considerations of CPU speed and computer memory currently 
available on scientific workstations typically
limit the number
of dipoles employed to $N < 10^6$ (see \S\ref{sec:memory_requirements}
for limitations on $N$ due to available RAM); 
for a given $N$, the limitations on $|m|kd$ 
translate into limitations on the ratio of target size to wavelength.

\noindent
For calculations of total cross sections $C_{\rm abs}$ and $C_{\rm sca}$,
we require $|m|kd < 1$:
\begin{equation}
a_{\rm eff} < 9.88 {\lambda\over |m|}\left({N\over10^6}\right)^{1/3}
{\rm ~~or~~} x < {62.04\over|m|}\left({N\over10^6}\right)^{1/3} ~~~.
\end{equation}
For scattering phase function calculations, we require $|m|kd < 0.5$:
\begin{equation}
a_{\rm eff} < 4.94 {\lambda\over |m|}\left({N\over10^6}\right)^{1/3}
{\rm ~~or~~} x < {31.02\over|m|}\left({N\over10^6}\right)^{1/3} ~~~.
\end{equation}

It is therefore clear that the DDA is not suitable for very large values of
the size parameter 
$x$, or very large values of the refractive index $m$.
The primary utility of the DDA is for scattering by dielectric 
targets with sizes comparable to the wavelength.
As discussed by Draine \& Goodman (1993), Draine \& Flatau (1994), and
Draine (2000),
total cross sections calculated with the DDA are 
accurate to a few percent provided
$N>10^4$ dipoles are used, criterion (\ref{eq:mkd_max}) is satisfied,
and $|m-1|< 2$.

Examples illustrating the accuracy of the DDA are shown in 
Figs.\ \ref{fig_Qm=1.33+0.01i}-\ref{fig_Qm=2+i} which show overall
scattering and absorption efficiencies as a function of wavelength for
different discrete dipole approximations to a sphere, with $N$ ranging
from 304 to 59728.
The DDA calculations assumed radiation incident along the (1,1,1)
direction in the ``target frame''.
Figs. \ {\ref{fig_dQdom=1.33+0.01i}-\ref{fig_dQdom=2+i} show the scattering
properties calculated with the DDA for $x=ka=7$.
Additional examples can be found in Draine \& Flatau (1994) and Draine (2000).

\begin{figure}
\centerline{\includegraphics[width=8.3cm]{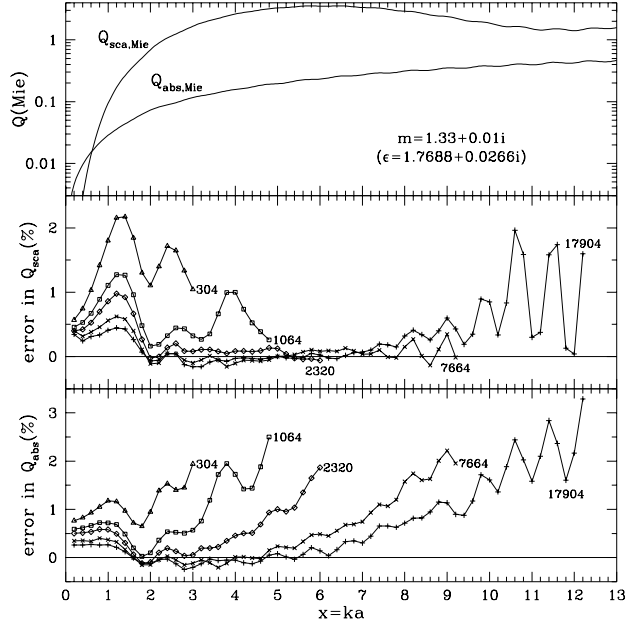}}
\caption{Scattering and absorption for a sphere with
	$m=1.33+0.01i$.  The upper panel shows the exact values of $Q_{\rm sca}$
	and $Q_{\rm abs}$, obtained with Mie theory, as functions of $x=ka$.
	The middle and lower panels show fractional errors in $Q_{\rm sca}$ and
	$Q_{\rm abs}$, obtained using {{\bf DDSCAT}}\ with polarizabilities 
	obtained
	from the Lattice Dispersion Relation, and labelled by the number $N$
	of dipoles in each pseudosphere.
	After Fig.\ 1 of Draine \& Flatau (1994).}
	\label{fig_Qm=1.33+0.01i}
\end{figure}

\begin{figure}
\centerline{\includegraphics[width=8.3cm]{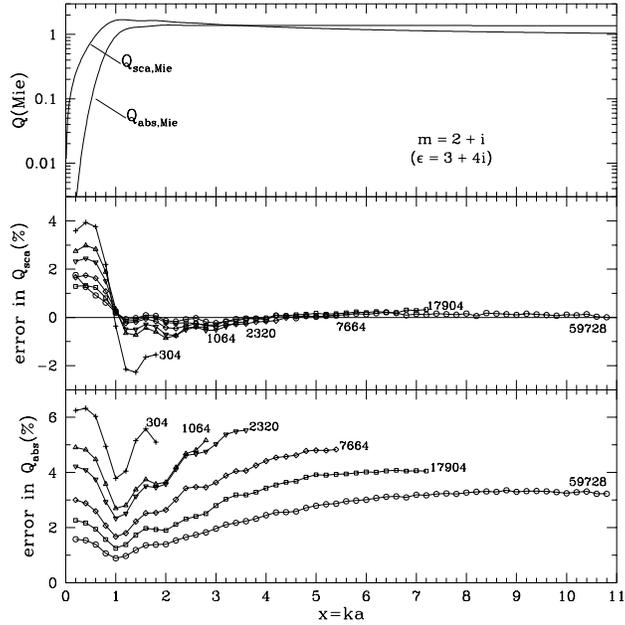}}
\caption{Same as Fig.\ \protect{\ref{fig_Qm=1.33+0.01i}},
	but for $m=2+i$. After Fig.\ 2 of Draine \& Flatau (1994).}
	\label{fig_Qm=2+i}
\end{figure}
\begin{figure}
\centerline{\includegraphics[width=8.3cm]{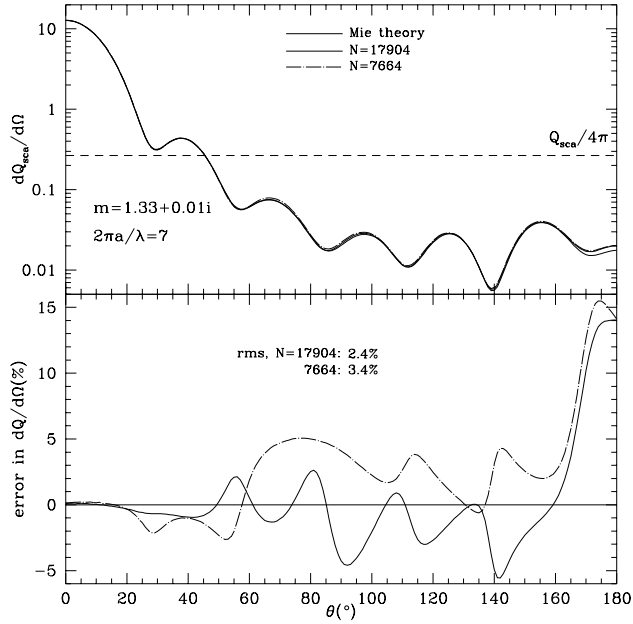}}
\caption{Differential scattering cross section for 
	$m=1.33+0.01i$ pseudosphere and $ka=7$.
	Lower panel shows fractional error compared to exact Mie theory
	result.
	The $N=17904$ pseudosphere has $|m|kd=0.57$, and an rms fractional
	error in $d\sigma/d\Omega$ of 2.4\%.
	After Fig.\ 5 of Draine \& Flatau (1994).}
	\label{fig_dQdom=1.33+0.01i}
\end{figure}
\begin{figure}
\centerline{\includegraphics[width=8.3cm]{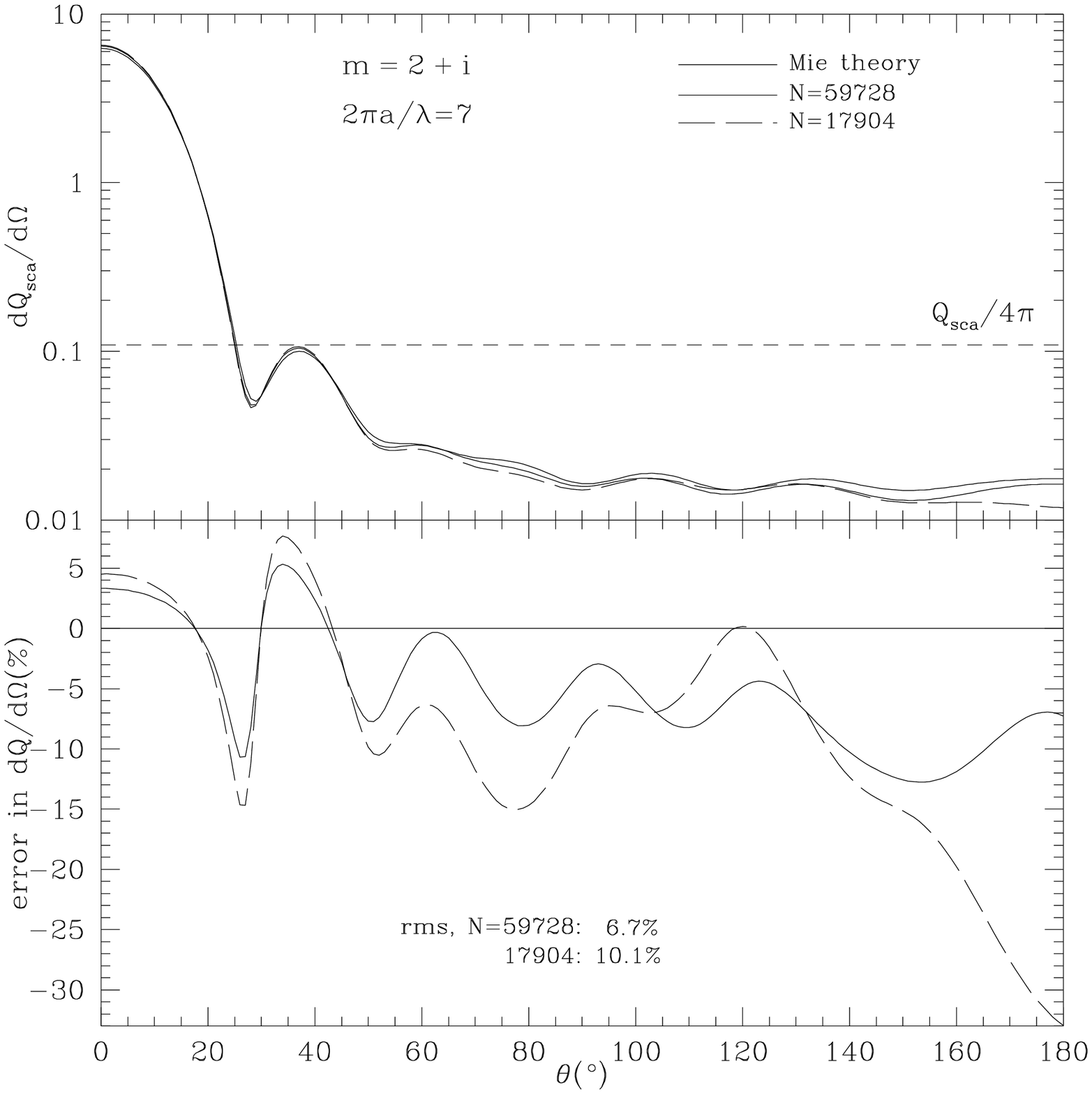}}
\caption{Same as Fig.\ \protect{\ref{fig_dQdom=1.33+0.01i}}
	but for $m=2+i$.
	The $N=59728$ pseudosphere has $|m|kd=0.65$, and an rms fractional
	error in $d\sigma/d\Omega$ of 6.7\%.
	After Fig.\ 8 of Draine \& Flatau (1994).}
	\label{fig_dQdom=2+i}
\end{figure}

\section{DDSCAT.5a\label{sec:DDSCATvers}}
\subsection{What Does It Calculate?}
{{\bf DDSCAT.5a}}\ solves the problem of scattering and absorption by an array of
polarizable point dipoles interacting with a monochromatic plane wave incident
from infinity.  
{{\bf DDSCAT.5a}}\ has the capability of automatically generating dipole
array representations for a variety of target geometries 
(see \S\ref{sec:target_generation}) and can also accept dipole
array representations of targets supplied by the user (although
the dipoles must be located on a cubic lattice).
The incident plane wave can have arbitrary elliptical
polarization (see \S\ref{sec:incident_polarization}), 
and the target can be arbitrarily oriented relative to the
incident radiation (see \S\ref{sec:target_orientation}).
The following quantities are calculated by {{\bf DDSCAT.5a}}\ :
\begin{itemize}
\item Absorption efficiency factor $Q_{\rm abs}\equiv C_{\rm abs}/\pi a_{\rm eff}^2$,
where $C_{\rm abs}$ is the absorption cross section;
\item Scattering efficiency factor $Q_{\rm sca}\equiv C_{\rm sca}/\pi a_{\rm eff}^2$,
where $C_{\rm sca}$ is the scattering cross section;
\item Extinction efficiency factor $Q_{\rm ext}\equiv Q_{\rm sca}+Q_{\rm abs}$;
\item Phase lag efficiency factor $Q_{\rm pha}$, defined so that the phase-lag
(in radians) of a plane wave after propagating a distance $L$ is just
$n_{t}Q_{\rm pha}\pi a_{\rm eff}^2 L$, 
where $n_{t}$ is the number density of targets.
\item The 4$\times$4 Mueller scattering intensity matrix $S_{ij}$ 
describing the complete
scattering properties of the target for scattering directions specified
by the user.
\item The radiation force efficiency vector ${\bf Q}_{pr}$ 
(see \S\ref{sec:torque_calculation}).
\item The radiation torque efficiency vector ${\bf Q}_\Gamma$
(see \S\ref{sec:torque_calculation}).
\end{itemize}

\subsection{Application to Targets in Dielectric Media
	\label{sec:target_in_medium}}
Let $\omega$ be the angular frequency of the incident radiation.
For many applications, the target is essentially {\it in vacuo}, in which
case the dielectric function $\epsilon$ which the user should supply
to {\tt DDSCAT} is the actual complex dielectric function 
$\epsilon_{\rm target}(\omega)$,
or complex refractive index 
$m_{\rm target}(\omega)=\sqrt{\epsilon_{\rm target}}$ 
of the target material.

However, for many applications of interest (e.g., marine optics, or
biological optics) the ``target'' body is embedded in a (nonabsorbing)
dielectric medium,
with (real) dielectric function $\epsilon_{\rm medium}(\omega)$, or
(real) refractive index $m_{\rm medium}(\omega)=\sqrt{\epsilon_{\rm medium}}$.
{{\bf DDSCAT.5a}} is fully applicable to these scattering problems, except
that:
\begin{itemize}
\item The ``dielectric function'' or ``refractive index'' 
supplied to {{\bf DDSCAT.5a}} should be
the {\it effective} dielectric function
\begin{equation}
\epsilon(\omega) = 
\frac{\epsilon_{\rm target}(\omega)}{\epsilon_{\rm medium}(\omega)}
\end{equation}
or {\it effective} refractive index:
\begin{equation}
m(\omega) =
\frac{m_{\rm target}(\omega)}{m_{\rm medium}(\omega)} .
\end{equation}
\item The wavelength $\lambda$ specified in {\tt ddscat.par} should be the
wavelength {\it in the medium}:
\begin{equation}
\lambda = \frac{\lambda_{vac}}{m_{\rm medium}},
\label{eq:lambda_medium}
\end{equation}
where $\lambda_{vac}=2\pi c/\omega$ is the wavelength {\it in vacuo}.
\end{itemize}
The absorption, scattering, extinction, and phase lag 
efficiency factors $Q_{\rm abs}$,
$Q_{\rm sca}$, and $Q_{\rm ext}$ calculated by {{\bf DDSCAT}}
will then be equal to the physical
cross sections for absorption, scattering, and extinction divided by
$\pi a_{\rm eff}^2$ -- e.g., the attenuation coefficient for radiation
propagating through a medium with a density $n_{t}$ of
scatterers will be just
$\alpha = n_{t}Q_{\rm ext}\pi a_{\rm eff}^2$.
Similarly, the phase lag (in radians) after propagating a distance $L$ will be
$n_{t}Q_{\rm pha}\pi a_{\rm eff}^2$.

The elements $S_{ij}$ of the 4$\times$4 Mueller scattering matrix ${\bf S}$
calculated by {{\bf DDSCAT}}
will be correct for scattering in the medium:
\begin{equation}
{\bf I}_{\rm sca} = 
\left(\frac{\lambda}{2\pi r}\right)^2 
{\bf S}\cdot {\bf I}_{\rm in} ,
\end{equation}
where ${\bf I}_{\rm in}$ and ${\bf I}_{\rm sca}$ are the Stokes vectors for
the incident and scattered light (in the medium), 
$r$ is the distance from the target,
and $\lambda$ is the wavelength in the medium (eq.\ \ref{eq:lambda_medium}).
See \S\ref{sec:mueller_matrix} for a detailed discussion of the
Mueller scattering matrix.

The time-averaged 
radiative force and torque (see \S\ref{sec:torque_calculation}) on a
target in a dielectric medium are
\begin{equation}
{\bf F}_{\rm rad} = {\bf Q}_{pr}\pi a_{\rm eff}^2 u_{\rm rad} ~~~,
\end{equation}
\begin{equation}
{\bf \Gamma}_{\rm rad} = 
{\bf Q}_\Gamma \pi a_{\rm eff}^2 u_{\rm rad} \frac{\lambda}{2\pi} ~~~,
\end{equation}
where the time-averaged energy density is
\begin{equation} 
u_{\rm rad}=\epsilon_{\rm medium} \frac{E_0^2}{8\pi} ~~~ ,
\end{equation}
where $E_0\cos(\omega t+\phi)$
is the electric field of the incident plane wave in the medium.

\bigskip
\section{What's New?\label{sec:whats_new}}
\subsection{{DDSCAT.5a}}
{{\bf DDSCAT.5a}}\ differs from
previous versions in five major respects:
\begin{enumerate}
\item Use of the new Generalized Prime Factor Algorithm (GPFA) 
developed by Clive Temperton (1992) for FFT calculations.  
The GPFA algorithm is generally faster than the previous algorithms, 
yet requires no more memory than the algorithm of Brenner (1969).
See \S\ref{sec:choice_of_fft}.
\item Availability of a new algorithm for iterative solution of
the system of complex linear equations.  
This algorithm is often faster than the algorithm of Petravic 
and Kuo-Petravic (1979) which was used through {{\bf DDSCAT}}{\bf .4b}
(and which remains available as an option in {{\bf DDSCAT.5a}}).
See \S\ref{sec:choice_of_algorithm}.
\item Automatic calculation of the transverse components of the radiative
force on the target.  See \S\ref{sec:torque_calculation}.
\item Capability to compute the electromagnetic torque on the target,
due to absorption and scattering of light from the incident beam,
as described by Draine and Weingartner (1996).
See \S\ref{sec:torque_calculation}
\item Automatic calculation of the 4$\times$4 Mueller scattering
matrix.
See \S\ref{sec:mueller_matrix}.
\item Improved output handling, including FORTRAN unformatted 
binary write option as well as a netCDF interface  and IDL postprocessing code.
\item Finally, this much-improved and expanded User Guide!
\end{enumerate}
We also call users' attention to a minor but possibly confusing change:
{{\bf DDSCAT.5a}}\ uses a different convention for specifying the target 
axes ${\hat{\bf a}}_1$ and ${\hat{\bf a}}_2$
for certain target choices ({\tt RCTNGL}, {\tt ELLIPS}, ...).  
We think the new convention is more straightforward.  See the discussion in 
\S\ref{sec:target_generation} below for details.

\subsection{{DSCAT.5a9}}
\noindent {{\bf DDSCAT.5a9}} differs from {\bf DDSCAT.5a8} in one
respect: 
\begin{itemize}
\item{{\bf DDSCAT.5a8}} contained an error in the
calculation of the elements of the scattering matrix
(see \S\ref{sec:mueller_matrix})
for scattering planes with $\phi\neq0$.
While {{\bf DDSCAT.5a8}} computed the scattering matrix 
correctly for $\phi_s=0$,
for irregular targets only the $S_{11}$ element was computed correctly
for scattering planes with $\phi_s\neq0$.
This bug has been corrected in {{\bf DDSCAT.5a9}}.
\end{itemize}

\subsection{{DDSCAT.5a10}}
\noindent {{\bf DDSCAT.5a10}} differs from {\bf DDSCAT.5a9} in two
respects:
\begin{itemize}
\item The public-domain LAPACK code has been replaced by the latest
routines obtained from NETLIB ({\tt http://www.netlib.org}).
Subroutine PRINAXIS did not function properly when the older LAPACK
routines were compiled using the {\tt g77} compiler, most likely due
to some variable not being {\tt SAVE}-d.
The new LAPACK code appears to execute properly.
\item A new shape option -- {\tt NSPHER} -- has been added, permitting
a target to be defined as the union of the volumes of an arbitrary
number of spheres.  The locations and radii of the {\tt NSPH} spheres
are input via a file.  Adding this option required the argument list
of subroutine {\tt TARGET} to be changed, and modifications were also
made to subroutines {\tt REAPAR} and {\tt REASHP}.
\item A new shape option -- {\tt PRISM3} -- has been added (on 2002.02.12),
enabling construction of a triangular prism target.  Concomitant modifications
were made to subroutines {\tt TARGET} and {\tt REAPAR}.
\end{itemize}

\section{Obtaining the Source Code\label{sec:downloading}}
The easiest way to download the source code is from the DDSCAT home page:\\
\hspace*{2em}{\tt http://www.astro.princeton.edu/$\sim$draine/DDSCAT.html}\\
where you
can obtain the file {\tt ddscat5a10.tar.gz} -- a ``gzipped tarfile''
containing the complete source code and documentation.
This can also be obtained by anonymous {\tt ftp} from
{\tt astro.princeton.edu}, sudirectory {\tt draine/scat/ddscat/ver5a10}.
Note that it is a compressed (binary) file.

The source code can be installed as follows.
First, place the file {\tt ddscat5a10.tar.gz}
in the directory where you would like {\bf DDSCAT}
to reside.  You should have at least 5 Mbytes of disk space available.

If you are on a Linux system, you should be able to type\\
\hspace*{2em}{\tt tar xvzf ddscat5a10.tar.gz}\\
which will ``gunzip'' the tarfile and then ``extract'' the files from the
tarfile.  If your version of ``tar'' doesn't support the ``z'' option
(e.g., you are running Solaris) then try\\
\hspace*{2em}{\tt zcat ddscat5a10.tar.gz | tar xvf -}\\
If neither of the above work on your system, try the two-stage procedure\\
\hspace*{2em}{\tt gunzip ddscat5a10.tar.gz}\\
\hspace*{2em}{\tt tar xvf ddscat5a10.tar}\\
(The disadvantage of the two-stage procedure is that it uses more disk
space, since after the second step you will have the uncompressed
tarfile {\tt ddscat5a10.tar} -- about 3.8 Mbytes --
in addition to all the files you have extracted
from the tarfile -- another 4.6 Mbytes).

Any of the above approaches should 
create subdirectories {\tt src, doc, misc,} and {\tt IDL}.
The source code will be in subdirectory {\tt src}, and documentation
in subdirectory {\tt doc}.

If you are running Windows on a PC, you will need the ``{\tt winzip}''
program, which can be downloaded from
{\tt http://www.winzip.com}.  {\tt winzip} should be able to ``unzip'' the
gzipped tarfile {\tt ddscat5a10.tar.gz} and ``extract'' the various files
from it automatically.

\section{Compiling and Linking\label{sec:compiling}}
In the discussion below, it will be assumed that the source 
code for {{\bf DDSCAT.5a}}\ has been installed in a directory {\tt DDA/src} .
To compile the code on a Unix system, position yourself in the 
directory {\tt DDA/src}.  
The {\tt Makefile} supplied has compiler options appropriate
for Sun Fortran under Solaris 2.x.  
If you have a different Fortran compiler, you will probably need 
to edit {\tt Makefile} to provide the desired compiler options.
{\tt Makefile} contains
samples of compiler options for selected operating systems, including 
HP AUX, IBM AIX, and SGI IRIX operating systems.

{\bf DDSCAT} can be compiled on Windows systems using standard Fortran
compilers (e.g., Compaq or Microsoft Visual Fortran), and run
successfully.  
Note, however,
that many of the instructions below are specifically for Unix or Linux
operating systems -- some steps may have to be done differently on a non-Unix
platform.

So far as we know, there are only two operating-system-dependent aspects of
{{\bf DDSCAT.5a}}: (1) the device number to use for ``standard output", and 
(2) the {\tt TIMEIT} routine.
There is, in addition, one installation-dependent aspect to the code: the
procedure for linking to the netCDF library (see \S\ref{subsec:netCDF} for
discussion of the possibility of writing binary files using the
machine-independent netCDF standard).

\subsection{Device Numbers {\tt IDVOUT} and {\tt IDVERR}
\label{subsec:IDVOUT}}
The variables {\tt IDVOUT} and {\tt IDVERR} specify device numbers 
for ``running output" and ``error messages", respectively.
Normally these would both be set to the device number
for ``standard output" (e.g., writing to the screen if running interactively).
The variables {\tt IDVOUT} and {\tt IDVERR} are 
set by {\tt DATA} statements in the ``main" program
{\tt DDSCAT.f} and in the output routine {\tt WRIMSG} (file {\tt wrimsg.f}).  
Under Sun Fortran, {\tt DATA IDVOUT/0/} 
results in unbuffered output to
``standard output"; unbuffered output (if available) is nice so that if you
choose to direct your output to a file (e.g., using 
{\tt ddscat >\& ddscat.out \&})
the output file will contain up-to-date information.  Other operating systems
or compilers may not support this, and you may need to edit {\tt DDSCAT.f} to
change the two data statements to {\tt DATA IDVOUT/6/} and 
{\tt DATA IDVERR/6/}, and edit {\tt wrimsg.f} to change {\tt DATA IDVOUT/0/}
to {\tt DATA IDVOUT/6/}.

\subsection{Subroutine {\tt TIMEIT}\label{subsec:timeit}}
The only other operating system-dependent part of {{\bf DDSCAT.5a}}\ is 
the single subroutine {\tt TIMEIT}.  
Several versions of {\tt TIMEIT} are provided:
\begin{itemize}
\item {\tt timeit\_sun.f} uses the SunOS system call {\tt etime}
\item {\tt timeit\_convex.f} uses the Convex OS system call {\tt etime}
\item {\tt timeit\_cray.f} uses the system call {\tt second}
\item {\tt timeit\_hp.f} uses the HP-AUX system calls {\tt sysconf} and 
{\tt times}
\item {\tt timeit\_ibm6000.f} uses the AIX system call {\tt mclock}
\item {\tt timeit\_osf.f} uses the DEC OSF system call {\tt etime}
\item {\tt timeit\_sgi.f} uses the IRIX system call {\tt etime}
\item {\tt timeit\_vms.f} uses the VMS system calls 
{\tt LIB\$INIT\_TIMER} and {\tt LIB\$SHOW\_TIMER}
\item {\tt timeit\_titan.f} uses the system call {\tt cputim}
\item {\tt timeit\_null.f} is a dummy routine which provides 
no timing information, but can be used under any operating system.
\end{itemize}
You {\it must} compile and link one of the 
{\tt timeit\_}{\it xxx}{\tt .f}
routines, possibly after modifying it to work on your system; 
{\tt timeit\_null.f}
is the easiest choice, but it will return no timing information.\footnote{
	Non-Unix sites: The VMS-compatible version of {\tt TIMEIT} is
	included in the file {\tt SRC9.FOR}.  For non-VMS sites, you will
	need to replace this version of {\tt TIMEIT} with one of the
	other versions, which are to be found in the file {\tt MISC.FOR}.
	If you are having trouble getting this to work, choose the
	``dummy'' version of {\tt TIMEIT} from {\tt MISC.FOR}: this will
	return no timing information, but is platform-independent.
	}

\subsection{Leaving the netCDF Capability Disabled}

The {\tt Makefile} supplied with the distribution of {{\bf DDSCAT.5a}}\ is
set up to link to a ``dummy'' subroutine {\tt dummywritenet.f} instead
of subroutine {\tt writenet.f}, in order to minimize problems during
initial compilation and linking.
The ``dummy'' routine has no functionality, other than bailing out
with a fatal error message if the user makes the mistake of trying
to specify one of the netCDF options ({\tt ALLCDF} or {\tt ORICDF}).
First-time users of {{\bf DDSCAT.5a}}\ should {\it not} try to use the
netCDF option -- simply skip this section, and specify option
{\tt NOTCDF} in {\tt ddscat.par}.
After successfully using {{\bf DDSCAT.5a}}\, you can return to 
\S\ref{subsec:enabling_netCDF}
to try to enable the netCDF capability.

\subsection{Enabling the netCDF Capability\label{subsec:enabling_netCDF}}

Subroutine {\tt WRITENET} (file {\tt writenet.f}) provides the capability
to output 
binary data in the netCDF standard format (see \S\ref{subsec:netCDF}).
In order to use this routine (instead of {\tt dummywritenet.f}), 
it is necessary to take the following steps:\footnote{
	Non-UNIX sites:
	You need to replace the ``dummy'' version of {\tt SUBROUTINE WRITENET}
	in {\tt SRC1.FOR} with the version provided in {\tt MISC.FOR}.
	You will also need to consult your systems administrator to verify
	that the netCDF library has already been installed on your system,
	and to find out how to link to the library routines.
	}
\begin{enumerate}
\item Have netCDF already installed on your system (check with your
	system administrator).
\item Find out where {\tt netcdf.inc} is located and
	edit the {\tt include} statement in {\tt writenet.f} to specify the
	correct path to {\tt netcdf.inc}.
\item Find out where the {\tt libnetcdf.a} library is located, and edit
	the Makefile so that the variable {\tt LIBNETCDF} specifies the correct
	path to this library.
\item Edit the Makefile to ``comment out'' (with a {\tt\#} symbol in column 1) 
the line {\tt writenet = dummywritenet} and ``uncomment'' (remove the 
{\tt\#} symbol) the line
{\tt writenet = writenet} so that {\tt writenet.f} will be compiled
instead of the dummy routine {\tt dummywritenet.f}.
\end{enumerate}

\subsection{Compiling and Linking...}
After suitably editing the Makefile 
(while still positioned in {\tt DDA/src}) 
simply type\footnote{
	Non-Unix sites: see \S\ref{subsec:nonunix}
	}\hfill\break
\indent\indent{\tt make ddscat}\hfill\break
which should create an executable file {\tt DDA/src/ddscat} .
The resulting executable will {\it not} have netCDF capability, and
will contain timing instructions compatible with the Solaris 1.x or 2.x 
operating systems on Sun computers, as well as several other versions of Unix.
To add netCDF capability, see \S\ref{subsec:enabling_netCDF}.
To replace the Sun-compatible timing routine with another, see
\S\ref{subsec:timeit}.

\subsection{Installation on Non-Unix Systems\label{subsec:nonunix}}

{{\bf DDSCAT.5a}} is written in standard Fortran-77 plus the
{\tt DO ... ENDDO} extension which appears to be supported by all current
Fortran-77 compatible compilers.  
It is possible to run {{\bf DDSCAT}} on non-Unix systems.
If the Unix ``make'' utility is not available, here in brief is what needs to 
be accomplished:

All of the necessary Fortran code to compile and link {{\bf DDSCAT.5a}}
is included in the following files:
{\tt SRC0.FOR}, {\tt SRC1.FOR}, {\tt SRC2.FOR},
{\tt SRC3.FOR},
{\tt SRC4.FOR}, {\tt SRC5.FOR}, {\tt SRC6.FOR}, {\tt SRC7.FOR}, {\tt SRC8.FOR},
{\tt SRC9.FOR}, {\tt CGCOMMON.FOR},
{\tt GPFA.FOR}, {\tt LAPACK.FOR}, and {\tt PIM.FOR}.
There is an additional file {\tt MISC.FOR}, but this is not 
needed for ``basic'' use of the code (see below).

The main program {\tt DDSCAT} is found in {\tt SRC0.FOR}.  It calls a number
of subroutines, which are included in the other {\tt *.FOR} files.

Three of the subroutines exist in more than one version.
The ``default'' version of each is located in the file {\tt SRC1.FOR}:
\begin{itemize}
\item Select the version of {\tt SUBROUTINE WRITENET} from {\tt SRC1.FOR}
	(do not use the version in {\tt MISC.FOR}).  The resulting code
	will not support {\bf netCDF} capability.  (If {\bf netCDF}
	capability is required, you will need to install {\bf netCDF} libraries
	on your system --
	see \S\ref{subsec:enabling_netCDF}).
\item Select the version of {\tt SUBROUTINE C3DFFT} from {\tt SRC1.FOR}
	(do not use the version in {\tt MISC.FOR})
\item There is one version of {\tt SUBROUTINE TIMEIT} 
	included in {\tt SRC1.FOR}, and a number of additional versions 
	in {\tt MISC.FOR}.
Use the version from {\tt SRC1.FOR}, which begins

\begin{verbatim}
      SUBROUTINE TIMEIT(CMSGTM,DTIME)
C
C     timeit_null
C
C This version of timeit is a dummy which does not provide any
C timing information.
\end{verbatim}
	This version does not use any system calls, and therefore the
	code should compile and link without problems; however, when you
	run the code, it will not report any timing information reporting
	how much time was spent on different parts of the calculation.

	If you wish to obtain
	timing information, you will need to find out what system calls
	are supported by your operating system.  You can look at the
	other versions of {\tt SUBROUTINE TIMEIT} in {\tt MISC.FOR} 
	to see how this has
	been done under VMS and various version of Unix.
\end{itemize}
Once you have selected the appropriate versions of {\tt WRITENET},
	{\tt C3DFFT}, and {\tt TIMEIT}, you
	can simply compile and link just as you would with any other Fortran
code with a number of modules.
You should end up with an executable with a (system-dependent) name
like {\tt DDSCAT.EXE}.

In addition to program {{\tt DDSCAT}}, we provide two other Fortran 77 
programs which may be useful.  Program {\tt CALLTARGET} can be used
to call the target generation routines which may be helpful for testing
purposes.
Program {\tt TSTFFT} is useful for comparing speeds of different FFT options.

\section{Moving the Executable}

Now reposition yourself into the directory {\tt DDA}\
(e.g., type {\tt cd ..}), 
and copy the executable
from {\tt src/ddscat} to the {\tt DDA}\ directory by typing

\indent\indent {\tt cp src/ddscat ddscat}

\noindent This should copy the file {\tt DDA/src/ddscat} to 
{\tt DDA/ddscat} .  Similarly,
copy the sample parameter file {\tt ddscat.par} and the file 
{\tt diel.tab} to the
{\tt DDA} directory by typing

\indent\indent{\tt cp doc/ddscat.par ddscat.par}\hfill\break
\indent\indent{\tt cp doc/diel.tab diel.tab}

\section{The Parameter File {\tt ddscat.par}\label{sec:parameter_file}}

The directory {\tt DDA}\ should now contain a sample file 
{\tt ddscat.par} which
provides parameters to the program {\tt ddscat}.  
As provided (see Appendix\ref{app:ddscat.par}),
the file {\tt ddscat.par} 
is set up to calculate scattering by a 8$\times$6$\times$4 
rectangular array of 192
dipoles, with an effective radius $a_{\rm eff}=1{\mu{\rm m}}$, at a wavelength of 
$6.2832{\mu{\rm m}}$ (for a ``size parameter'' $2\pi a_{\rm eff}/\lambda=1$).

The dielectric function of the target material is provided in the file
{\tt diel.tab}, which is a sample file in which the refractive index is set to
$m=1.33+0.01i$ at all wavelengths; the name of this file is provided to 
{\tt ddscat} by
the parameter file {\tt ddscat.par}.

The sample parameter file as supplied calls 
for the new GPFA FFT routine ({\tt GPFAFT}) 
of Temperton (1992) to be employed and the {\tt PBCGST} iterative method
to be used for solving the system of linear equations.
(See section \S\ref{sec:choice_of_fft} and \S\ref{sec:choice_of_algorithm} 
for discussion of choice of FFT algorithm and 
choice of equation-solving algorithm.)

The sample parameter file specifies (via option {\tt LATTDR}) that the 
``Lattice Dispersion Relation'' of Draine and Goodman (1993) 
be employed to determine the
dipole polarizabilities.
See \S\ref{sec:polarizabilities} for discussion of other options.

The sample {\tt ddscat.par} file specifies that the calculations be done for a
single wavelength ($6.2832{\mu{\rm m}}$) and a single effective radius 
($1.00{\mu{\rm m}}$).
Note that in {{\bf DDSCAT.5a}}\ the ``effective radius'' 
$a_{\rm eff}$ is the radius of a sphere of
equal volume -- i.e., a sphere of volume $Nd^3$ , where $d$ 
is the lattice spacing
and $N$ is the number of occupied (i.e., non-vacuum) 
lattice sites in the target.
Thus the effective radius $a_{\rm eff} = (3N/4\pi)^{1/3}d$ .

The incident radiation is always assumed to propagate along the $x$ axis in
the ``Lab Frame''.  
The sample {\tt ddscat.par} file specifies incident polarization
state ${\hat{\bf e}}_{01}$ to be along the $y$ axis 
(and consequently polarization state ${\hat{\bf e}}_{02}$
will automatically be taken to be along the $z$ axis).  
{\tt IORTH=2} in {\tt ddscat.par}
calls for calculations to be carried out for both incident polarization
states (${\hat{\bf e}}_{01}$ and ${\hat{\bf e}}_{02}$
-- see \ref{sec:incident_polarization}).

The target is assumed to have two vectors ${\hat{\bf a}}_1$ and ${\hat{\bf a}}_2$ 
embedded in it; ${\hat{\bf a}}_2$ is perpendicular to ${\hat{\bf a}}_1$.  
In the case of the 8$\times$6$\times$4 rectangular array of the
sample calculation, the vector ${\hat{\bf a}}_1$ is along the ``long'' axis of 
the target, and the vector ${\hat{\bf a}}_2$ is along the ``intermediate'' axis.  
The target orientation in the Lab Frame is set by three angles: $\beta$,
$\Theta$, and $\Phi$, defined and discussed below in 
\S\ref{sec:target_orientation}.  
Briefly, the polar angles
$\Theta$ and $\Phi$ specify the direction of ${\hat{\bf a}}_1$ in the Lab Frame.  
The target is assumed to be rotated around ${\hat{\bf a}}_1$ by an angle $\beta$.  
The sample {\tt ddscat.par} file
specifies $\beta=0$ and $\Phi=0$ (see lines in {\tt ddscat.par}
specifying variables {\tt BETA} and {\tt PHI}), 
and calls for three values of the angle
$\Theta$ (see line in {\tt ddscat.par} specifying variable {\tt THETA}).  
{{\bf DDSCAT.5a}}\ chooses $\Theta$ values uniformly spaced in
$\cos\Theta$; thus, asking for three values of $\Theta$ between 
0 and $90^\circ$ yields $\Theta=0$, $60^\circ$, and $90^\circ$.

Appendix \ref{app:ddscat.par} provides a detailed description of the 
file {\tt ddscat.par}.\footnote{
	Note that the structure of {\tt ddscat.par} depends on the version 
	of {{\bf DDSCAT}}\ being used.
	Make sure you update old parameter files before using them with
	{{\bf DDSCAT.5a}}\ !
	}

\section{Running {{\bf DDSCAT.5a}}\ Using the Sample {\tt ddscat.par} File}

To execute the program on a UNIX system (running either {\tt sh} 
or {\tt csh}), simply type

\indent\indent {\tt ddscat >\& ddscat.out \&}

\noindent which will redirect the ``standard output'' to the file {\tt ddscat.out}, 
and run the calculation in the background.  
The sample calculation (8x6x4=192 dipole 
target, 3 orientations, two incident polarizations, with scattering calculated 
for 14 distinct scattering directions), requires 1.0 cpu seconds on a Sun 
Ultra-170 workstation.

\section{Output Files}

\subsection{ASCII files\label{subsec:ascii}}
If you run DDSCAT using the command\hfill\break
\indent\indent{\tt ddscat >\& ddscat.out \&} \hfill\break
you will have various types of ASCII files when the computation is complete:
\begin{itemize}
\item a file {\tt ddscat.out};
\item a file {\tt mtable};
\item a file {\tt qtable};
\item a file {\tt qtable2};
\item files {\tt w}{\it xx}{\tt r}{\it yy}{\tt ori.avg} 
(one, {\tt w00r00ori.avg}, for the sample calculation);
\item if {\tt ddscat.par} specified {\tt IWRKSC}=1, there will also be 
files {\tt w}{\it xx}{\tt r}{\it yy}{\tt k}{\it zzz}{\tt .sca} 
(3 for the sample calculation: {\tt w00r00k000.sca},
{\tt w00r00k001.sca}, {\tt w00r00k002.sca}).
\end{itemize}
The file {\tt ddscat.out} will contain any error messages generated as well as
a running report on the progress of the calculation, including creation of the
target dipole array.  During the iterative calculations, $Q_{\rm ext}$, 
$Q_{\rm abs}$, and $Q_{\rm pha}$ are printed after each iteration; you will be able 
to judge the degree to which convergence has been achieved.  
Unless {\tt TIMEIT} has been disabled, there will also be timing information.

The file {\tt mtable} contains a summary of the dielectric constant used in
the calculations.

The file {\tt qtable} contains a summary of the orientationally-averaged values
of $Q_{\rm ext}$, $Q_{\rm abs}$, $Q_{\rm sca}$, 
$g(1)=\langle\cos(\theta_s)\rangle$, and $Q_{bk}$.  
Here $Q_{\rm ext}$, $Q_{\rm abs}$, and $Q_{\rm sca}$ are the extinction, absorption, 
and scattering cross sections divided by
$\pi a_{\rm eff}^2$.  
$Q_{bk}$ is the differential cross section for backscattering (area per sr)
divided by $\pi a_{\rm eff}^2$.

The file {\tt qtable2} contains a summary of the orientationally-averaged 
values of $Q_{\rm pha}$, $Q_{\rm pol}$, and $Q_{\rm cpol}$.  
Here $Q_{\rm pha}$ is the ``phase shift'' cross section
divided by $\pi a_{\rm eff}^2$ (see definition in Draine 1988).  
$Q_{\rm pol}$ is the ``polarization efficiency factor'', 
equal to the difference between $Q_{\rm ext}$ for the two
orthogonal polarization states.
We define a ``circular polarization efficiency factor''
$Q_{\rm cpol}\equiv Q_{\rm pol}Q_{\rm pha}$, 
since an optically-thin medium with a small twist in the alignment
direction will produce circular polarization in initially unpolarized light in
proportion to $Q_{\rm cpol}$.

For each wavelength and size, {{\bf DDSCAT.5a}}\ produces a 
file with a name of 
the form\break{\tt w{\it xx}r{\it yy}ori.avg}, where index 
{\it xx} (=00, 01, 02....) 
designates the wavelength and index {\it yy} (=00, 01, 02...) designates the 
``radius''; this file contains $Q$ values
and scattering information averaged over however many target orientations
have been specified (see \S\ref{sec:target_orientation}.  
The file {\tt w00r00ori.avg}
produced by the sample calculation is provided below in 
Appendix \ref{app:w00r00ori.avg}.

   In addition, if {\tt ddscat.par} has specified {\tt IWRKSC}=1 
(as for the sample calculation), {{\bf DDSCAT.5a}}\ will generate files with 
names of the form {\tt w{\it xx}r{\it yy}k{\it zzz}.avg}, 
where {\it xx} and {\it yy} are as before, and 
index {\it zzz} =(000,001,002...)
designates the target orientation; these files contain $Q$ values and 
scattering information for {\it each} of the target orientations.  
The structure of each of these files is very similar to that of the 
{\tt w{\it xx}r{\it yy}ori.avg} files.  
Because these files may not be of particular interest, and take up disk 
space, you may choose to set {\tt IWRKSC}=0 in future work.  
However, it is suggested that you run the sample calculation with 
{\tt IWRKSC}=1.

   The sample {\tt ddscat.par} file specifies {\tt IWRKSC}=1 and calls for 
use of 1 wavelength, 1 target size, and averaging over 3 target orientations.
Running {{\bf DDSCAT.5a}}\ with the sample {\tt ddscat.par} file will therefore 
generate files {\tt w00r00k000.sca}, {\tt w00r00k001.sca}, and 
{\tt w00r00k002.sca} .
To understand the information contained in one of these files, please consult
Appendix \ref{app:w00r00k000.sca}, which contains an example of 
the file {\tt w00r00k000.sca} produced in the sample calculation.

\subsection{Binary Option\label{subsec:binary}}

It is possible to output an ``unformatted'' or ``binary'' file ({\tt dd.bin})
with fairly complete information, including header and data sections.
This is accomplished by specifying either {\tt ALLBIN} or {\tt ORIBIN} in
{\tt ddscat.par} .

Subroutine {\tt writebin.f} provides an example of how this can be done.
The ``header'' section contains dimensioning 
and other variables which do not change with 
wavelength, particle geometry, and target orientation.
The header section contains
data defining the particle shape, wavelengths, particle
sizes, and target orientations.
If {\tt ALLBIN} has been specified, the ``data'' section contains,
for each orientation,
Mueller matrix results for each scattering direction.
The data output is limited to actual dimensions of arrays;
e.g. {\tt nscat,4,4} elements  of Mueller matrix are written
rather than {\tt mxscat,4,4}. 
This is an important consideration
when writing postprocessing codes. 

A skeletal example of a postprocessing code was written by us 
and is provided in subdirectory {\tt DDA/IDL}. 
If you do plan to use the Interactive Data Language (IDL) for postprocessing,
you may consider using the netCDF binary file option which offers
substantial advantages over the FORTRAN unformatted write.
More information about IDL is available at
{\verb|http://www.rsinc.com/idl|}. 
Unfortunately IDL requires a license and hence is not 
distributed with {{\bf DDSCAT}}. 

\subsection{Machine-Independent Binary File Option: netCDF
	\label{subsec:netCDF}}

The ``unformatted'' binary file is compact, fairly complete, and 
(in comparison to ASCII output files) is efficiently
written from FORTRAN. 
However, binary files
are not compatible between different computer architectures if written
by ``unformatted'' {\tt WRITE} from FORTRAN.
Thus, you have to process the data on the same computer architecture
that was used for the {{\bf DDSCAT}}\ calculations.
We provide an option of using netCDF with DDSCAT.
The netCDF library defines a
machine-independent format for representing scientific data. Together, 
the interface, library, and format support the creation, access, 
and sharing of scientific data. 
For more information see {\tt http://www.unidata.ucar.edu/packages/netcdf}.
 
Several major graphics packages (for example IDL) have 
adopted netCDF as a standard for data transport. 
For this reason, and because of strong and free support 
of netCDF over the network by UNIDATA, we have implemented a netCDF 
interface in {{\bf DDSCAT}}. 
This may become the method of choice for binary file storage of
output from {{\bf DDSCAT}}.

After the initial ``learning curve'' netCDF offers many advantages: 
\begin{itemize}
\item It is easy to examine the contents of the file (using tools such as 
{\tt ncdump}).
\item Binary files are portable - they  can be created on a supercomputer and 
processed on a workstation.
\item Major graphics packages now offer netCDF interfaces.
\item Data input and output are an order of magnitude faster than for ASCII 
files.
\item Output data files are compact.
\end{itemize}
The disadvantages include: 
\begin{itemize}
\item Need to have netCDF installed on your system.
\item Lack of support of complex numbers.
\item Nontrivial learning curve for using netCDF.
\item Lack of portability of netCDF libraries.
\end{itemize}

The public-domain netCDF 
software is available for many operating systems
from \break{\tt http://www.unidata.ucar.edu/packages/netcdf} .
The steps necessary for enabling the netCDF capability in {{\bf DDSCAT.5a}}\
are listed above in \S\ref{subsec:enabling_netCDF}.
Once these have been successfully accomplished, and you are ready to
run {{\bf DDSCAT}}\ to produce netCDF output, 
you must choose either the {\tt ALLCDF} or {\tt ORICDF} option in 
{\tt ddscat.par}; {\tt ALLCDF} will result in a file which will
include the Mueller matrix for every wavelength, particle size, 
and orientation,
whereas {\tt ORICDF} will result in a file limited to just the
orientational averages for each wavelength and target size.

\section{Dipole Polarizabilities\label{sec:polarizabilities}}

Option {\tt LATTDR} specifies that the 
``Lattice Dispersion Relation'' of Draine and Goodman (1993) 
be employed to determine the
dipole polarizabilities.; other possible choices are {\tt DRAI88} (prescription
used by Draine 1988) and {\tt GOBR88} 
(prescription used by Goedecke \& O'Brien 1988
and Hage \& Greenberg 1990).
In the limit $|m|kd \ll 1$ (where $k=2\pi/\lambda$ is the wave
vector and $d$ is the lattice spacing) all three options converge to the same
limit; for $|m|kd > 0.1$ 
there are significant differences among them.  Consult
the paper by Draine \& Goodman (1993) for 
discussion and comparison of these
three prescriptions.  
{\bf Option {\tt LATTDR} is recommended for general use},
based upon the tests presented by Draine \& Goodman (1993).

\section{Dielectric Functions\label{sec:dielectric_func}}

In order to assign the appropriate dipole polarizabilities, {{\bf DDSCAT.5a}}\ must be
given the dielectric constant of the material (or materials) of which the
target of interest is composed. 
Unless the user wishes to use the dielectric function of either solid or
liquid H$_2$O (see below),
his information is supplied to {{\bf DDSCAT}}\ through a table (or tables), 
read by subroutine {\tt DIELEC} in file {\tt dielec.f}, and providing
either the complex refractive index $m=n+ik$ or complex dielectric
function $\epsilon=\epsilon_1+i\epsilon_2$ as a function of wavelength
$\lambda$.
Since $m=\epsilon^{1/2}$, or $\epsilon=m^2$, 
the user must supply either $m$ or $\epsilon$,
but not both.

The table formatting is intended to be quite flexible.
The first line of the table consists of text, up to 80 characters of 
which will be read and included in the output to identify the choice of 
dielectric function.
(For the sample problem, it consists of simply the statement
{\tt m = 1.33 + 0.01i}.)
The second line consists of 5 integers; either the second and third {\it or}
the fourth and fifth should be zero.
\begin{itemize}
\item The first integer specifies which column the wavelength is stored in.
\item The second integer specifies which column Re$(m)$ is stored in.
\item The third integer specifies which column Im$(m)$ is stored in.
\item The fourth integer specifies which column Re$(\epsilon)$ is stored in.
\item The fifth integer specifies which column Im$(\epsilon)$ is stored in.
\end{itemize}
If the second and third integers are zeros, then {\tt DIELEC} will read
Re$(\epsilon)$ and Im$(\epsilon)$ from the file;
if the fourth and fifth integers are zeros, then Re$(m)$ and Im$(m)$ will
be read from the file.

The third line of the file is used for column headers, and the data begins
in line 4.
{\it There must be at least 3 lines of data:} even if 
$\epsilon$ is required at only one wavelength, please supply 
two additional ``dummy'' wavelength entries in the table so 
that the interpolation apparatus will not be confused.

In the event that the user wishes to compute scattering by targets with
the refractive index of either solid or liquid H$_2$O, we have included
two ``built-in'' dielectric functions.
If {\tt H2OICE} is specified on line 10 of {\tt ddscat.par}, {{\bf DDSCAT}}\
will use the dielectric function of H$_2$O ice at T=250K as compiled by 
Warren (1984).
If {\tt H2OLIQ} is specified on line 10 of {\tt ddscat.par}, {{\bf DDSCAT}}\
will use the dielectric function for liquid H$_2$O at T=280K 
using subroutine {\tt REFWAT}
written by Eric A. Smith.
For more information see 
{\tt http://atol.ucsd.edu/$\sim$pflatau/scatlib/refr.html} .

\section{Choice of FFT Algorithm\label{sec:choice_of_fft}}
One major change in going from {{\bf DDSCAT}}{\bf .4b} to {\bf 4c} and {\bf 5a} 
was modification of 
the code to permit use of the GPFA FFT algorithm developed by Dr. Clive 
Temperton.  DDSCAT continues to offer both the Brenner code as well as the 
``old'' Temperton code as alternative FFT implementations.  
The ``old'' Temperton 
code requires about 11\% more memory than either the Brenner or GPFA codes 
(to use the ``old'' Temperton algorithm, 
{\tt DDSCAT.f} must be compiled with {\tt MXMEM}=1 rather than {\tt MXMEM}=0).

To help persuade the user that the GPFA code is an important step forward, 
we provide a program {\tt TSTFFT} to compare the performance of 
different 3-D FFT implementations.  
To compile, link, and run this program on a Unix system,\footnote{ 
	Non-Unix systems: the {\tt TSTFFT} Fortran source 
	code is in the file {\tt MISC.FOR}.}
position yourself in the {\tt DDA/src} directory and type

\indent\indent {\tt make tstfft}\hfill\break
\indent\indent {\tt tstfft}

\noindent Output results will be written into a file {\tt res.dat}.  
Here is a copy
of the {\tt res.dat} file created when run on a Sun Ultrasparc 170:
\begin{verbatim}
 CPU time (sec) for different 3-D FFT methods
 Machine= Sun Ultrasparc-170
 parameter LVR = 64
 LVR="length of vector registers" in gpfa2f,gpfa3f,gpfa5f
==========================================================
                     Brenner      Temperton      Temperton
   NX  NY  NZ        (FOURX)        (Old)          (GPFA)
    2   2   2       0.000068       0.000065       0.000262
    3   3   3       0.000200       0.000077       0.000096
    4   4   4       0.000064       0.000086       0.000111
    5   5   5       0.000651       0.000292       0.000137
    6   6   6       0.001171       0.000170       0.000223
    8   8   8       0.000350       0.000250       0.000323
    9   9   9       0.004654       0.000300       0.000505
   10  10  10       0.005218       0.000457       0.000575
   12  12  12       0.008664       0.000814       0.000751
   15  15  15       0.023387       0.001583       0.001798
   16  16  16       0.003137       0.001359       0.002386
   18  18  18       0.042196       0.003908       0.003479
   20  20  20       0.043754       0.003881       0.004010
   24  24  24       0.075672       0.010722       0.007198
   25  25  25       0.103492       0.008142       0.011232
   27  27  27       0.160173       0.011126       0.012293
   30  30  30       0.185177       0.017111       0.014631
   32  32  32       0.042211       0.034032       0.023245
   36  36  36       0.322581       0.071013       0.027502
   40  40  40       0.379551       0.133391       0.044182
   45  45  45       0.797107       0.199537       0.069740
   48  48  48       0.668408       0.255849       0.094520
   50  50  50       0.983722       0.287208       0.115833
   54  54  54       1.530319       0.427839       0.142388
   60  60  60       1.804154       0.607189       0.181221
   64  64  64       1.005013       0.806311       0.342762
   72  72  72       3.812469       1.219364       0.362328
   75  75  75       5.364067       1.234528       0.437496
   80  80  80       4.509048       1.576118       0.563839
   81  81  81       8.017456       1.798195       0.560326
   90  90  90      10.201881       2.455399       0.779869
\end{verbatim}

It is clear that the GPFA code is generally {\it much} faster than the 
Brenner FFT
(by a factor of 13 for the 90$\times$90$\times$90 case) 
and significantly faster than the 
``old'' Temperton FFT code (by a factor of 3 for the 90$\times$90$\times$90 
case), although for some cases the differences are not large (e.g., 
27$\times$27$\times$27).  
Since the GPFA code is memory-efficient as well, it appears to be the method 
of choice on scalar machines.  It appears to also be best on vector machines.

The GPFA code contains a parameter {\tt LVR} which is set in 
{\tt data} statements in the
routines {\tt gpfa2f}, {\tt gpfa3f}, and {\tt gpfa5f}.  
{\tt LVR} is supposed to be optimized to correspond 
to the ``length of a vector register'' on vector machines.  
As delivered, this parameter is set to 64, which is supposed to be 
appropriate for Crays other than the C90 (for the C90, 128 is supposed to 
be preferable).\footnote{
	In place of ``preferable'' users are encouraged to 
	read ``necessary'' -- there is some basis for fearing that
	results computed on a C90 with {\tt LVR} other than 128 run the
	risk of being incorrect!  Please use {\tt LVR=128} if running
	on a C90!}
The value of {\tt LVR} is not critical for scalar machines, as long as 
it is fairly large.  We found little difference between {\tt LVR}=64 and 
128 on a Sparc 10/51 and on an Ultrasparc 170.  You may wish to 
experiment with different {\tt LVR} values on your computer architecture.  
To change {\tt LVR}, you need to edit {\tt gpfa.f} and change the three 
{\tt data} statements where {\tt LVR} is set.

The choice of FFT implementation is obtained by specifying one of:
\begin{itemize}
\item{\tt GPFAFT} to use the GPFA algorithm (Temperton 1992) -- {\bf this is
	recommended!};
\item{\tt TMPRTN} to obtain the ``old'' Temperton (1983) implementation;
\item{\tt BRENNR} to obtain the {\tt FOURX} implementation of 
	the Brenner (1969) algorithm;
\item{\tt CONVEX} to use the native FFT routine on a Convex system
\end{itemize}

\section{Choice of Iterative Algorithm\label{sec:choice_of_algorithm}}

As discussed elsewhere (e.g., Draine 1988), the problem of electromagnetic
scattering of an incident wave ${\bf E}_{inc}$ by an array of $N$ point dipoles
can be
cast in the form
\begin{equation}
{\bf A} {\bf P} = {\bf E}
\label{eq:AP=E}
\end{equation}
where ${\bf E}$ is a $3N$-dimensional (complex) 
vector of the incident electric field ${\bf E}_{inc}$ at the
$N$ lattice sites, ${\bf P}$ is a $3N$-dimensional (complex) vector of the (unknown)
dipole
polarizations, and ${\bf A}$ is a $3N\times3N$ complex matrix.

Because $3N$ is a large number, direct methods for solving this system of equations
for the unknown vector ${\bf P}$ are impractical, but iterative methods are
useful: we begin with a guess (typically, ${\bf P}=0$) for the unknown polarization
vector, and then iteratively improve the estimate for ${\bf P}$ until 
equation (\ref{eq:AP=E}) is solved to some error criterion.
The error tolerance may be
specified as
\begin{equation}
{|{\bf A}^\dagger {\bf A} {\bf P} - {\bf A}^\dagger {\bf E}| \over
| {\bf A}^\dagger {\bf E} |}
 < h 
\label{eq:err_tol}~~~,
\end{equation}
where ${\bf A}^\dagger$ is the Hermitian conjugate of ${\bf A}$
[$(A^\dagger)_{ij} \equiv (A_{ji})^*$], and
$h$ is the error tolerance.
We typically use $h=10^{-5}$ in order to satisfy eq.(\ref{eq:AP=E}) to high accuracy.
The error tolerance $h$ can be specified by the user 
(see Appendix \ref{app:ddscat.par}).

A major change in going from {{\bf DDSCAT}}{\bf .4b} to {\bf 5a} is the 
implementation of
several different algorithms for iterative solution of the system of complex
linear equations.  {{\bf DDSCAT.5a}}\ has been modified to permit solution algorithms
to be treated in a fairly ``modular'' fashion, facilitating the testing of
different algorithms.  
Many algorithms were compared by Flatau (1997)\footnote{
	A postscript copy of this report -- file {\tt cg.ps} -- is
	distributed with the {{\bf DDSCAT.5a}}\ documentation.}; 
two of them performed well
and are made available to the user in {{\bf DDSCAT.5a}}.
The choice of algorithm is made by specifying one of the options:
\begin{itemize}
\item {\tt PBCGST} -- Preconditioned BiConjugate Gradient with 
	STabilitization method from the Parallel Iterative Methods 
	(PIM) package created by R. Dias da Cunha and T. Hopkins.
\item {\tt PETRKP} -- the complex conjugate gradient algorithm of 
	Petravic \& Kuo-Petravic (1979), as coded in the Complex Conjugate 
	Gradient package (CCGPACK) created by P.J. Flatau.
	This is the algorithm discussed by Draine (1988) and used in 
	previous versions of {{\bf DDSCAT}}.
\end{itemize}
Both methods work well.  
Our experience suggests that {\tt PBCGST} is often
faster than {\tt PETRKP}, by perhaps a factor of two.  
We therefore recommend it, but include the {\tt PETRKP} method 
as an alternative.

The Parallel Iterative Methods (PIM) by Rudnei Dias da Cunha 
({\tt rdd@ukc.ac.uk}) and Tim Hopkins ({\tt trh@ukc.ac.uk}) is a collection of 
Fortran 77 routines designed to solve systems of
linear equations on parallel and scalar computers using a variety 
of iterative methods (available at \hfill\break
{\tt http://www.mat.ufrgs.br/pim-e.html}).
PIM offers a number of iterative methods, including 
\begin{itemize}
\item     Conjugate-Gradients, CG (Hestenes 1952), 
\item     Bi-Conjugate-Gradients, BICG (Fletcher 1976), 
\item     Conjugate-Gradients squared, CGS (Sonneveld 1989), 
\item     the stabilised version of Bi-Conjugate-Gradients, BICGSTAB (van der Vorst 1991), 
\item     the restarted version of BICGSTAB, RBICGSTAB (Sleijpen \& Fokkema 1992) 
\item     the restarted generalized minimal residual, RGMRES (Saad 1986), 
\item     the restarted generalized conjugate residual, RGCR (Eisenstat 1983), 
\item     the normal equation solvers, CGNR (Hestenes 1952 and CGNE (Craig 1955), 
\item     the quasi-minimal residual, QMR (highly parallel version due to 
	Bucker \& Sauren 1996), 
\item     transpose-free quasi-minimal residual, TFQMR (Freund 1992), 
\item     the Chebyshev acceleration, CHEBYSHEV (Young 1981). 
\end{itemize}
The source code for these methods is distributed with {\tt DDSCAT} but
only {\tt PBCGST} and {\tt PETRKP} can be called directly via 
{\tt ddscat.par}. It is possible (and was done)
to add other options by changing the code in {\tt getfml.f} .
A helpful introduction to conjugate gradient methods is provided by the report
``Conjugate Gradient Method Without Agonizing Pain" by Jonathan R. Shewchuk,
available as a postscript file:
{\tt ftp://REPORTS.ADM.CS.CMU.EDU/usr0/anon/1994/CMU-CS-94-125.ps}.

\section{Calculation of Radiative Force and Torque\label{sec:torque_calculation}}
In addition to solving the scattering problem for a dipole array, {{\bf DDSCAT.5a}}\
can compute the three-dimensional force ${\bf F}_{\rm rad}$ 
and torque ${\bf \Gamma}_{\rm rad}$ exerted on this array by
the incident and scattered radiation fields.  This calculation is carried out,
after solving the scattering problem, provided {\tt DOTORQ} has been
specified in {\tt ddscat.par}.
For each incident polarization mode, 
the results are given in terms of dimensionless efficiency vectors
${\bf Q}_{pr}$ and ${\bf Q}_{\Gamma}$, defined by
\begin{equation}
{\bf Q}_{pr} \equiv {{\bf F}_{\rm rad} \over 
\pi a_{\rm eff}^2 u_{\rm rad}} ~~~,
\end{equation}
\begin{equation}
{\bf Q}_\Gamma \equiv {k{\bf \Gamma}_{\rm rad} \over
\pi a_{\rm eff}^2 u_{\rm rad}} ~~~,
\end{equation}
where ${\bf F}_{\rm rad}$ and ${\bf \Gamma}_{\rm rad}$ are the time-averaged
force and torque on the dipole array,
$k=2\pi/\lambda$ is the wavenumber {\it in vacuo},
and $u_{\rm rad} = E_0^2/8\pi$ 
is the time-averaged energy density for
an incident plane wave with amplitude 
$E_0 \cos(\omega t + \phi)$.
The radiation pressure efficiency vector can be written
\begin{equation}
{\bf Q}_{pr} = Q_{\rm ext}{\hat{\bf k}} - Q_{\rm sca}{\bf g} ~~~,
\end{equation}
where ${\hat{\bf k}}$ is the direction of propagation of the incident radiation,
and the vector {\bf g} is the mean direction of propagation of the
scattered radiation:
\begin{equation}
{\bf g} = {1\over C_{\rm sca}}\int d\Omega
{dC_{\rm sca}({\hat{\bf n}},{\hat{\bf k}})\over d\Omega} {\hat{\bf n}} ~~~,
\end{equation}
where $d\Omega$ is the element of solid angle in scattering direction ${\hat{\bf n}}$,
and $dC_{\rm sca}/d\Omega$ is the differential scattering cross section.

Equations for the evaluation of the radiative force and torque
are derived by Draine \& Weingartner (1996).
It is important to note that evaluation of ${\bf Q}_{pr}$ and
${\bf Q}_\Gamma$ involves averaging over scattering directions to
evaluate the linear and angular momentum transport by the scattered
wave.
This evaluation requires appropriate choices of the parameters
{\tt ICTHM} and {\tt IPHM} -- see \S\ref{sec:averaging_scattering}.

\section{Memory Requirements \label{sec:memory_requirements}}
Since Fortran-77 does not allow for dynamic memory allocation, 
the executable has compiled into it the dimensions for a number of arrays;
these array dimensions limit the size of the dipole array which the code can
handle.  
The source code supplied to you (which can be used to run the sample
calculation with 192 dipoles) is restricted to problems with targets which
are maximum extent of 8 lattice spacings along the $x$-, $y$-, and 
$z$-directions ({\tt MXNX=8,MXNY=8,MXNZ=8}; i.e, the target must fit within 
an 8$\times$8$\times$8=512 cube) and involve at most 9 different dielectric 
functions ({\tt MXCOMP=9}).  
With this dimensioning, the executable requires about 1.3 MB of memory to run on an
Ultrasparc system; 
memory requirements on other hardware and with other compilers
should be similar.  
To run larger problems, you will need to edit the code to
change {\tt PARAMETER} values and recompile.

All of the dimensioning takes place in the main program {\tt DDSCAT} -- this
should be the only routine which it is necessary to recompile.  All of the
dimensional parameters are set in {\tt PARAMETER} statements appearing before
the array declarations.  You need simply edit the parameter statements.
Remember, of course, that the amount of memory allocated by the code when it
runs will depend upon these dimensioning parameters, so do not set them to
ridiculously large values!  The parameters {\tt MXNX}, {\tt MXNY}, 
{\tt MXNZ} specify the maximum extent of the target in the $x$-, $y$-, or 
$z$-directions.  
Set the parameter {\tt MXMEM=0} or {\tt 1} depending on whether workspace 
required by Temperton's FFT algorithm is to be reserved.  
The parameter {\tt MXCOMP} specifies the maximum number
of different dielectric functions which the code can handle at one time.  
The comment statements in the code supply all the information you should 
need to change these parameters.

The memory requirement for {{\bf DDSCAT.5a}}\ (with the netCDF option disabled) is 
approximately 
$(1059+0.623{\tt MXNX}\times{\tt MXNY}\times{\tt MXNZ})$ kbytes
when compiled with {\tt MXMEM=0}, or 
$(1059+0.686{\tt MXNX}\times{\tt MXNY}\times{\tt MXNZ})$ kbytes 
when compiled with {\tt MXMEM=1}.  
Thus with {\tt MXMEM=0} a 32$\times$32$\times$32 calculation requires 
21.0 MBytes,
while a 48$\times$48$\times$48 calculation requires 68.4 MBytes.
These values are for an Ultrasparc system using the Sun FORTRAN compiler.

\section{Target Orientation \label{sec:target_orientation}}

Recall that we define a ``Lab Frame'' (LF) in which the incident radiation
propagates in the $+x$ direction.  
In {\tt ddscat.par} one specifies the first
polarization state ${\hat{\bf e}}_{01}$ 
(which obviously must lie in the $y,z$ plane in the LF);
{{\bf DDSCAT}}\ automatically constructs a second polarization state 
${\hat{\bf e}}_{02} = {\hat{\bf x}} \times {\hat{\bf e}}_{01}^*$
orthogonal to ${\hat{\bf e}}_{01}$ (here ${\hat{\bf x}}$ is the unit vector 
in the $+x$ direction of the LF.
For purposes of discussion we will
always let unit vectors ${\hat{\bf x}}$, ${\hat{\bf y}}$, 
${\hat{\bf z}}={\hat{\bf x}}\times{\hat{\bf y}}$ be the three coordinate axes of the LF.
Users will often find it convenient to let polarization 
vectors ${\hat{\bf e}}_{01}={\hat{\bf y}}$,
${\hat{\bf e}}_{02}={\hat{\bf z}}$ (although this is not mandatory -- 
see \S\ref{sec:incident_polarization}).

% this figure seems to end up on a page by itself, so let is be
% full-size (16cm) rather than half-size (8 cm)
\begin{figure}
\centerline{\includegraphics[width=16.cm]{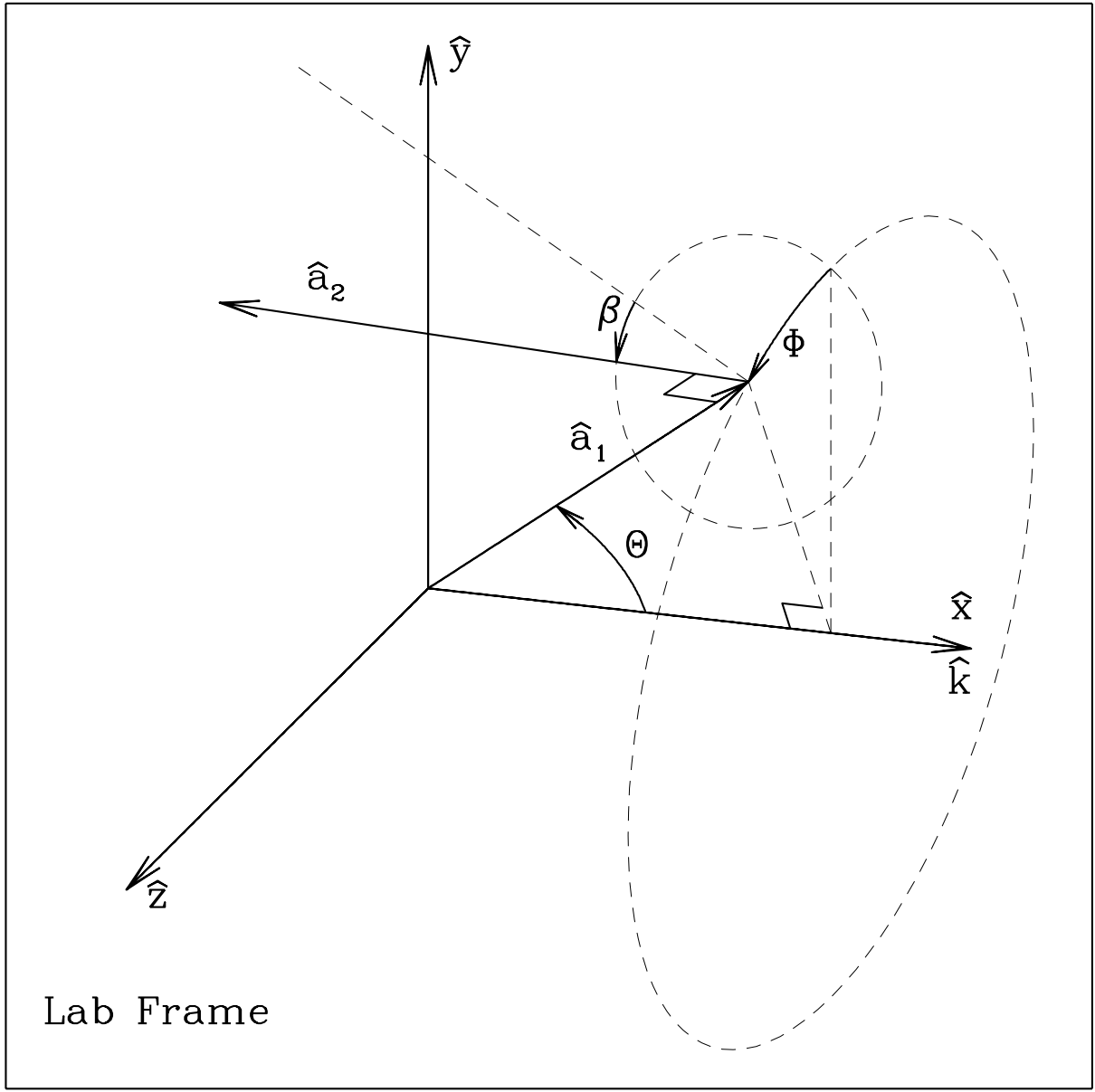}}
\caption{Target orientation in the Lab Frame.  ${\hat{\bf x}}$ is
	the direction of propagation of the incident radiation, and ${\hat{\bf y}}$ is
	the direction of the ``real'' component of the
	first incident polarization mode.
	In this coordinate system, the orientation of target axis ${\hat{\bf a}}_1$
	is specified by angles $\Theta$ and $\Phi$.
	With target axis ${\hat{\bf a}}_1$ fixed, the orientation of target axis 
	${\hat{\bf a}}_2$ is then determined by angle $\beta$ specifying rotation of 
	the target around ${\hat{\bf a}}_1$.  
	When $\beta=0$, ${\hat{\bf a}}_2$ lies in the ${\hat{\bf a}}_1$,${\hat{\bf x}}$
	plane.}
	\label{fig_target_orientation}
\end{figure}

Recall that definition of a target involves specifying two unit vectors,
${\hat{\bf a}}_1$ and ${\hat{\bf a}}_2$, which are imagined to be ``frozen'' into the target.  
We require ${\hat{\bf a}}_2$ to be orthogonal to ${\hat{\bf a}}_1$.  
Therefore we may define a ``Target Frame" (TF) defined by the three unit
vectors ${\hat{\bf a}}_1$, ${\hat{\bf a}}_2$, and ${\hat{\bf a}}_3 = {\hat{\bf a}}_1 \times {\hat{\bf a}}_2$ .

For example, when {{\bf DDSCAT}}\ creates a 8$\times$6$\times$4 
rectangular solid, it fixes ${\hat{\bf a}}_1$ to be
along the longest dimension of the solid, and ${\hat{\bf a}}_2$ to be 
along the next-longest dimension.

Orientation of the target relative to the incident radiation can in
principle be determined two ways:
\begin{enumerate}
\item specifying the direction of ${\hat{\bf a}}_1$ and ${\hat{\bf a}}_2$ in the LF, or
\item specifying the directions of ${\hat{\bf x}}$ (incidence direction) and 
${\hat{\bf y}}$ in the TF.
\end{enumerate}
{{\bf DDSCAT.5a}}\ uses method 1.: the angles $\Theta$, $\Phi$, and 
$\beta$ are specified in the file {\tt ddscat.par}.  
The target is oriented such that the polar angles
$\Theta$ and $\Phi$ specify the direction of ${\hat{\bf a}}_1$ 
relative to the incident direction ${\hat{\bf x}}$, where
the ${\hat{\bf x}}$,${\hat{\bf y}}$ plane has $\Phi=0$.  
Once the direction of ${\hat{\bf a}}_1$ is specified, the
angle $\beta$ than specifies how the target is to rotated 
around the axis ${\hat{\bf a}}_1$ to fully specify its orientation.  
A more extended and precise explanation follows:

\subsection{Orientation of the Target in the Lab Frame}

DDSCAT uses three angles, $\Theta$, $\Phi$, and $\beta$, 
to specify the directions
of unit vectors ${\hat{\bf a}}_1$ and ${\hat{\bf a}}_2$ in the LF
(see Fig.\ \ref{fig_target_orientation}).

$\Theta$ is the angle between ${\hat{\bf a}}_1$ and ${\hat{\bf x}}$.

When $\Phi=0$, ${\hat{\bf a}}_1$ will lie in the ${\hat{\bf x}},{\hat{\bf y}}$ plane.  
When $\Phi$ is nonzero, it will
refer to the rotation of ${\hat{\bf a}}_1$ around ${\hat{\bf x}}$: e.g., 
$\Phi=90^\circ$ puts ${\hat{\bf a}}_1$ in the ${\hat{\bf x}},{\hat{\bf z}}$ plane.

When $\beta=0$, ${\hat{\bf a}}_2$ will lie in the ${\hat{\bf x}},{\hat{\bf a}}_1$ plane, 
in such a way that when $\Theta=0$ and $\Phi=0$, ${\hat{\bf a}}_2$ is in the
${\hat{\bf y}}$ direction: e.g, $\Theta=90^\circ$, $\Phi=0$, $\beta=0$ has 
${\hat{\bf a}}_1={\hat{\bf y}}$ and
${\hat{\bf a}}_2=-{\hat{\bf x}}$.  
Nonzero $\beta$ introduces an additional rotation of ${\hat{\bf a}}_2$ around
${\hat{\bf a}}_1$: e.g., $\Theta=90^\circ$, $\Phi=0$, $\beta=90^\circ$ 
has ${\hat{\bf a}}_1={\hat{\bf y}}$ and ${\hat{\bf a}}_2={\hat{\bf z}}$.

Mathematically:
\begin{eqnarray}
{\hat{\bf a}}_1 &=&   {\hat{\bf x}} \cos\Theta 
+ {\hat{\bf y}} \sin\Theta \cos\Phi 
+ {\hat{\bf z}} \sin\Theta \sin\Phi
	\\
{\hat{\bf a}}_2 &=& - {\hat{\bf x}} \sin\Theta \cos\beta 
+ {\hat{\bf y}} [\cos\Theta \cos\beta \cos\Phi-\sin\beta \sin\Phi] \nonumber\\
&&+ {\hat{\bf z}} [\cos\Theta \cos\beta \sin\Phi+\sin\beta \cos\Phi]
	\\
{\hat{\bf a}}_3 &=&   {\hat{\bf x}} \sin\Theta \sin\beta 
- {\hat{\bf y}} [\cos\Theta \sin\beta \cos\Phi+\cos\beta \sin\Phi] \nonumber\\
  &&         - {\hat{\bf z}} [\cos\Theta \sin\beta \sin\Phi-\cos\beta \cos\Phi]
\end{eqnarray}
or, equivalently:
\begin{eqnarray}
{\hat{\bf x}} &=&   {\hat{\bf a}}_1 \cos\Theta
           - {\hat{\bf a}}_2 \sin\Theta \cos\beta
           + {\hat{\bf a}}_3 \sin\Theta \sin\beta \\
{\hat{\bf y}} &=&   {\hat{\bf a}}_1 \sin\Theta \cos\Phi
           + {\hat{\bf a}}_2 [\cos\Theta \cos\beta \cos\Phi-\sin\beta \sin\Phi]
\nonumber\\
&&           - {\hat{\bf a}}_3 [\cos\Theta \sin\beta \cos\Phi+\cos\beta \sin\Phi]
\\
{\hat{\bf z}} &=&   {\hat{\bf a}}_1 \sin\Theta \sin\Phi
           + {\hat{\bf a}}_2 [\cos\Theta \cos\beta \sin\Phi+\sin\beta \cos\Phi]
\nonumber\\
&&           - {\hat{\bf a}}_3 [\cos\Theta \sin\beta \sin\Phi-\cos\beta \cos\Phi]
\end{eqnarray}

\subsection{Orientation of the Incident Beam in the Target Frame}
   Under some circumstances, one may wish to specify the target orientation
such that ${\hat{\bf x}}$ (the direction of propagation of the radiation) 
and ${\hat{\bf y}}$ (usually the first polarization direction) 
and ${\hat{\bf z}}$ (= ${\hat{\bf x}} \times {\hat{\bf y}})$ refer to certain directions
in the TF.  Given the definitions of the LF and TF above, this is simply an
exercise in coordinate transformation.  
For example, one might wish to have
the incident radiation propagating along the (1,1,1) direction in the TF
(example 14 below).  Here we provide some selected examples:
\begin{enumerate}
\item${\hat{\bf x}}= {\hat{\bf a}}_1$, ${\hat{\bf y}}= {\hat{\bf a}}_2$, ${\hat{\bf z}}= {\hat{\bf a}}_3$ : 
	$\Theta=  0$, $\Phi+\beta=  0$
\item${\hat{\bf x}}= {\hat{\bf a}}_1$, ${\hat{\bf y}}= {\hat{\bf a}}_3$, ${\hat{\bf z}}=-{\hat{\bf a}}_2$ : 
	$\Theta=  0$, $\Phi+\beta= 90^\circ$
\item${\hat{\bf x}}= {\hat{\bf a}}_2$, ${\hat{\bf y}}= {\hat{\bf a}}_1$, ${\hat{\bf z}}=-{\hat{\bf a}}_3$ : 
	$\Theta= 90^\circ$, $\beta=180^\circ$, $\Phi=  0$
\item${\hat{\bf x}}= {\hat{\bf a}}_2$, ${\hat{\bf y}}= {\hat{\bf a}}_3$, ${\hat{\bf z}}= {\hat{\bf a}}_1$ : 
	$\Theta= 90^\circ$, $\beta=180^\circ$, $\Phi= 90^\circ$
\item${\hat{\bf x}}= {\hat{\bf a}}_3$, ${\hat{\bf y}}= {\hat{\bf a}}_1$, ${\hat{\bf z}}= {\hat{\bf a}}_2$ : 
	$\Theta= 90^\circ$, $\beta=-90^\circ$, $\Phi=  0$
\item${\hat{\bf x}}= {\hat{\bf a}}_3$, ${\hat{\bf y}}= {\hat{\bf a}}_2$, ${\hat{\bf z}}=-{\hat{\bf a}}_1$ : 
	$\Theta= 90^\circ$, $\beta=-90^\circ$, $\Phi=-90^\circ$
\item${\hat{\bf x}}=-{\hat{\bf a}}_1$, ${\hat{\bf y}}= {\hat{\bf a}}_2$, ${\hat{\bf z}}=-{\hat{\bf a}}_3$ : 
	$\Theta=180^\circ$, $\beta+\Phi=180^\circ$
\item${\hat{\bf x}}=-{\hat{\bf a}}_1$, ${\hat{\bf y}}= {\hat{\bf a}}_3$, ${\hat{\bf z}}= {\hat{\bf a}}_2$ : 
	$\Theta=180^\circ$, $\beta+\Phi= 90^\circ$
\item${\hat{\bf x}}=-{\hat{\bf a}}_2$, ${\hat{\bf y}}= {\hat{\bf a}}_1$, ${\hat{\bf z}}= {\hat{\bf a}}_3$ : 
	$\Theta= 90^\circ$, $\beta=  0$, $\Phi=  0$
\item${\hat{\bf x}}=-{\hat{\bf a}}_2$, ${\hat{\bf y}}= {\hat{\bf a}}_3$, ${\hat{\bf z}}=-{\hat{\bf a}}_1$ : 
	$\Theta= 90^\circ$, $\beta=  0$, $\Phi=-90^\circ$
\item${\hat{\bf x}}=-{\hat{\bf a}}_3$, ${\hat{\bf y}}= {\hat{\bf a}}_1$, ${\hat{\bf z}}=-{\hat{\bf a}}_2$ : 
	$\Theta= 90^\circ$, $\beta=-90^\circ$, $\Phi=  0$
\item${\hat{\bf x}}=-{\hat{\bf a}}_3$, ${\hat{\bf y}}= {\hat{\bf a}}_2$, ${\hat{\bf z}}= {\hat{\bf a}}_1$ : 
	$\Theta= 90^\circ$, $\beta=-90^\circ$, $\Phi= 90^\circ$
\item${\hat{\bf x}}=({\hat{\bf a}}_1+{\hat{\bf a}}_2)/\surd2$, ${\hat{\bf y}}={\hat{\bf a}}_3$, 
	${\hat{\bf z}}=({\hat{\bf a}}_1-{\hat{\bf a}}_2)/\surd2$: 
	$\Theta=45^\circ$, $\beta=180^\circ$, $\Phi=90^\circ$
\item${\hat{\bf x}}=({\hat{\bf a}}_1+{\hat{\bf a}}_2+{\hat{\bf a}}_3)/\surd3$, 
	${\hat{\bf y}}=({\hat{\bf a}}_1-{\hat{\bf a}}_2)/\surd2$, 
	${\hat{\bf z}}=({\hat{\bf a}}_1+{\hat{\bf a}}_2-2{\hat{\bf a}}_3)/\surd6$:
	$\Theta=54.7356^\circ$, $\beta=135^\circ$, $\Phi=30^\circ$.
\end{enumerate}
\subsection{Sampling in $\Theta$, $\Phi$, and $\beta$\label{subsec:sampling}}
The present version, {{\bf DDSCAT.5a}}, chooses the angles $\beta$, $\Theta$, 
and $\Phi$ to sample the intervals ({\tt BETAMI,BETAMX}), 
({\tt THETMI,THETMX)}, ({\tt PHIMIN,PHIMAX}), where
{\tt BETAMI}, {\tt BETAMX}, {\tt THETMI}, {\tt THETMX}, {\tt PHIMIN}, 
{\tt PHIMAX} are specified in {\tt ddscat.par} .
The prescription for choosing the angles is to:
\begin{itemize}
\item uniformly sample in $\beta$;
\item uniformly sample in $\Phi$;
\item uniformly sample in $\cos\Theta$.
\end{itemize}
This prescription is appropriate for random orientation of the target, within
the specified limits of $\beta$, $\Phi$, and $\Theta$.

Note that when {{\bf DDSCAT.5a}}\ chooses angles it handles $\beta$ 
and $\Phi$ differently
from $\Theta$.\footnote{
	This is a change from {{\bf DDSCAT}}{\bf .4a}!!.}  
The range for $\beta$ is
divided into {\tt NBETA} intervals, and the midpoint of each interval is 
taken.  
Thus, if you take {\tt BETAMI}=0, {\tt BETAMX}=90, {\tt NBETA}=2 you will get 
$\beta=22.5^\circ$ and $67.5^\circ$.
Similarly, if you take {\tt PHIMIN}=0, {\tt PHIMAX}=180, {\tt NPHI}=2 
you will get $\Phi=45^\circ$ and $135^\circ$.

   Sampling in $\Theta$ is done quite differently from sampling in $\beta$ 
and $\Phi$.
First, as already mentioned above, {{\bf DDSCAT.5a}}\ samples uniformly 
in $\cos\Theta$, not $\Theta$.  
Secondly, the sampling depends on whether {\tt NTHETA} is even or odd.
\begin{itemize}
\item If {\tt NTHETA} is odd, then the values of $\Theta$ selected include the 
extreme values {\tt THETMI} and {\tt THETMX}; thus, {\tt THETMI}=0, 
{\tt THETMX}=90, {\tt NTHETA}=3 will give you $\Theta=0,60^\circ,90^\circ$.  
\item If {\tt NTHETA} is even, then the range of
$\cos\Theta$ will be divided into {\tt NTHETA} intervals, 
and the midpoint of each interval will be taken; thus, {\tt THETMI}=0,
{\tt THETMX}=90, {\tt NTHETA}=2 will give you
$\Theta=41.41^\circ$ and $75.52^\circ$ [$\cos\Theta=0.25$ and $0.75$].  
\end{itemize}
The reason for this is that
if odd {\tt NTHETA} is specified, then the ``integration'' over $\cos\Theta$
is performed using Simpson's rule for greater accuracy.  
If even {\tt NTHETA} is specified, then
the integration over $\cos\Theta$ is performed by simply taking the 
average of the results for the different $\Theta$ values.

If averaging over orientations is desired, it is recommended that the
user specify an {\it odd} value of {\tt NTHETA} so that Simpson's rule will 
be employed.

\section{Orientational Averaging\label{sec:orientational_averaging}}

{{\bf DDSCAT}} has been constructed to facilitate the computation of orientational
averages.  How to go about this depends on the distribution of orientations
which is applicable.

\subsection{Randomly-Oriented Targets}

For randomly-oriented targets, we wish to compute the orientational average of
a quantity $Q(\beta,\Theta,\Phi)$:
\begin{equation}
\langle Q \rangle = {1\over 8\pi^2}\int_0^{2\pi}d\beta
\int_{-1}^1 d\cos\Theta
\int_0^{2\pi}d\Phi ~ Q(\beta,\Theta,\Phi) ~~~.
\end{equation}
To compute such averages, all you need to do is edit the file
{\tt ddscat.par}
so that DDSCAT knows what ranges of the angles $\beta$, $\Theta$, 
and $\Phi$ are of interest.  
For a randomly-oriented target with no symmetry, you would need to
let $\beta$ run from 0 to $360^\circ$, 
$\Theta$ from 0 to $180^\circ$, 
and $\Phi$ from 0 to $360^\circ$.

For targets with symmetry, on the other hand, the ranges of $\beta$, $\Theta$,
and $\Phi$ may be reduced.  
First of all, remember that averaging over $\Phi$ is relatively ``inexpensive",
so when in doubt average over 0 to $360^\circ$; 
most of the computational ``cost" is
associated with the number of different values of ($\beta$,$\Theta$) which 
are used.
Consider a cube, for example, with axis ${\hat{\bf a}}_1$ normal to one of the cube 
faces; for this cube $\beta$ need run only from 0 to $90^\circ$, since the 
cube has fourfold symmetry
for rotations around the axis ${\hat{\bf a}}_1$.  
Furthermore, the angle $\Theta$ need run only
from 0 to $90^\circ$, since the orientation 
($\beta$,$\Theta$,$\Phi$) is indistinguishable from
($\beta$, $180^\circ-\Theta$, $360^\circ-\Phi$).

For targets with symmetry, the user is encouraged to test the significance
of $\beta$,$\Theta$,$\Phi$ on targets with small numbers of dipoles (say, of 
the order of 100 or so) but having the desired symmetry.

It is important to remember that {{\bf DDSCAT}}{\bf .4b} handled even and odd values of
{\tt NTHETA} differently -- see \S\ref{sec:parameter_file} above!  
For averaging over random
orientations odd values of {\tt NTHETA} 
are to be preferred, as this will allow use
of Simpson's rule in evaluating the ``integral" over $\cos\Theta$.

\subsection{Nonrandomly-Oriented Targets}

Some special cases (where the target orientation distribution is
uniform for rotations around the $x$ axis = direction of propagation of the
incident radiation), one may be able to use {{\bf DDSCAT.5a}} \
with appropriate choices of input parameters.  
More generally, however, you will need to modify
subroutine {\tt ORIENT} to generate a list of {\tt NBETA} values of $\beta$, 
{\tt NTHETA} values of $\Theta$, and 
{\tt NPHI} values of $\Phi$, plus two weighting arrays 
{\tt WGTA(1-NTHETA,1-NPHI)}
and {\tt WGTB(1-NBETA)}.  
Here {\tt WGTA} gives the weights which should be attached
to each ($\Theta$,$\Phi$) orientation, and {\tt WGTB} gives the weight to be 
attached to each $\beta$ orientation.  
Thus each orientation of the target is to be weighted
by the factor {\tt WGTA}$\times${\tt WGTB}.   
For the case of random orientations, {{\bf DDSCAT.5a}}\ chooses
$\Theta$ values which are uniformly spaced in $\cos\Theta$, and 
$\beta$ and $\Phi$ values which are uniformly spaced, and therefore uses 
uniform weights\hfill\break
\indent\indent {\tt WGTB}=1./{\tt NBETA}\hfill\break
When {\tt NTHETA} is even, {{\bf DDSCAT}}\ sets\hfill\break
\indent\indent {\tt WGTA}=1./({\tt NTHETA}$\times${\tt NPHI})\hfill\break
but when {\tt NTHETA} is odd, {{\bf DDSCAT}}\ uses Simpson's rule when integrating 
over $\Theta$ and\hfill\break
\indent\indent {\tt WGTA}= (1/3 or 4/3 or 2/3)/({\tt NTHETA}$\times${\tt NPHI})

Note that the program structure of {{\bf DDSCAT}}\ may not be ideally suited for
certain highly oriented cases.  If, for example, the orientation is such
that for a given $\Phi$ value only one $\Theta$ value is possible (this 
situation might describe ice needles oriented with the long axis 
perpendicular to the vertical in the Earth's atmosphere, illuminated by the 
Sun at other than the zenith) then it is foolish to consider all the 
combinations of $\Theta$ and
$\Phi$ which the present version of {{\bf DDSCAT}}\ is set up to do.  
We hope to improve this in a future version of {{\bf DDSCAT}}.

\section{Target Generation\label{sec:target_generation}}

DDSCAT contains routines to generate dipole 
arrays representing targets of various
geometries, including spheres, ellipsoids, rectangular solids, cylinders,
hexagonal prisms, tetrahedra, two touching ellipsoids, and three touching
ellipsoids.  
The target type is determined by the variable
{\tt CSHAPE} (= {\tt ELLIPS}, {\tt CYLNDR}, etc., ...) on line 7 of
{\tt ddscat.par}, and up to 6 parameters ({\tt SHPAR1}, {\tt SHPAR2}, 
{\tt SHPAR3}, ...) on line 8.
The target geometry is most conveniently described in a coordinate
system attached to the target which we refer to as the ``Target Frame'' 
(TF), so in this section {\it only} we will let x,y,z be coordinates in
the Target Frame.
Once the target is generated, the orientation of the target in the
Lab Frame is accomplished as described in \S\ref{sec:target_orientation}.

Target geometries currently supported include:
\begin{itemize}
\item {\tt ELLIPS} -- {\bf Homogeneous, isotropic ellipsoid.}\\
	``Lengths'' 
	{\tt SHPAR1}, {\tt SHPAR2},
	{\tt SHPAR3} in the $x$, $y$, $z$ directions in the TF.
	$(x/{\tt SHPAR1})^2+(y/{\tt SHPAR2})^2+(z/{\tt SHPAR3})^2 = d^2/4$,
	where $d$ is the interdipole spacing.\\
	The target axes are set to ${\hat{\bf a}}_1=(1,0,0)$ and 
	${\hat{\bf a}}_2=(0,1,0)$ in the TF.\\
	User must set {\tt NCOMP}=1 on line 9 of {\tt ddscat.par}.\\
	A {\bf homogeneous, isotropic sphere} is obtained by setting 
	{\tt SHPAR1} = {\tt SHPAR2} = {\tt SHPAR3} = diameter/$d$.
\item {\tt ANIELL} -- {\bf Homogeneous, anisotropic ellipsoid.}\\
	{\tt SHPAR1}, {\tt SHPAR2},
	{\tt SHPAR3} have same meaning as for {\tt ELLIPS}.
	Target axes ${\hat{\bf a}}_1=(1,0,0)$ and 
	${\hat{\bf a}}_2=(0,1,0)$ in the TF. 
	It is assumed that the dielectric tensor is diagonal in the TF. 
	User must set {\tt NCOMP}=3 and provide $xx$, $yy$, $zz$ elements of
	the dielectric tensor.
\item {\tt CONELL} -- {\bf Two concentric ellipsoids.}\\
	{\tt SHPAR1}, {\tt SHPAR2},
	{\tt SHPAR3} specify the lengths along $x$, $y$, $z$ axes (in the TF)
	of the {\it outer} ellipsoid; {\tt SHPAR4}, {\tt SHPAR5}, {\tt SHPAR6} 
	are the lengths, along the $x$, $y$, $z$ axes (in the TF) of the
	{\it inner} ellipsoid.  
	The ``core" within the inner ellipsoid is composed of isotropic 
	material 1; 
	the ``mantle" between inner and outer
	ellipsoids is composed of isotropic material 2.  
	Target axes ${\hat{\bf a}}_1=(1,0,0)$, ${\hat{\bf a}}_2=(0,1,0)$ 
	in TF.  
	User must set {\tt NCOMP}=2 and provide dielectric functions for 
	``core'' and ``mantle'' materials.
\item {\tt CYLNDR} -- {\bf Homogeneous, isotropic cylinder.}\\
	Length/$d$={\tt SHPAR1}, 
	diameter/$d$={\tt SHPAR2}, with cylinder axis 
	$={\hat{\bf a}}_1=(1,0,0)$
	and ${\hat{\bf a}}_2=(0,1,0)$ in the TF.
	User must set {\tt NCOMP}=1.
\item {\tt UNICYL} -- {\bf Homogeneous cylinder with unixial anisotropic 
	dielectric tensor.}\\
	{\tt SHPAR1}, {\tt SHPAR2} have same meaning as for
	{\tt CYLNDR}.
	Cylinder axis $={\hat{\bf a}}_1=(1,0,0)$, ${\hat{\bf a}}_2=(0,1,0)$. 
	It is assumed that the dielectric tensor $\epsilon$ 
	is diagonal in the TF,
	with $\epsilon_{yy}=\epsilon_{zz}$.
	User must set
	{\tt NCOMP}=2.
	Dielectric function 1 is for ${\bf E} \parallel {\bf\hat{a}}_1$
	(cylinder axis), dielectric function 2 is for 
	${\bf E} \perp {\bf\hat{a}}_1$.
\item {\tt RCTNGL} -- {\bf Homogeneous, isotropic, rectangular solid.}\\
	x, y, z lengths/$d$ = {\tt SHPAR1}, {\tt SHPAR2}, {\tt SHPAR3}.  
	Target axes ${\hat{\bf a}}_1=(1,0,0)$ and ${\hat{\bf a}}_2=(0,1,0)$ 
	in the TF.
	User must set {\tt NCOMP}=1.
\item {\tt HEXGON} -- {\bf Homogeneous, isotropic hexagonal prism.}\\
	{\tt SHPAR1}=Length/$d$, 
	{\tt SHPAR2} = 2$\times$(hexagon side)/$d$ =
	(distance between opposite vertices of hexagon)/$d$ = 
	``hexagon diameter''/$d$.
	Target axis ${\hat{\bf a}}_1 = (1,0,0)$ in the TF is along the
	prism axis, and target
	axis ${\hat{\bf a}}_2=(0,1,0)$ in the TF is
	normal to one of the rectangular faces
	of the hexagonal prism.
	User must set {\tt NCOMP}=1.
\item {\tt TETRAH} -- {\bf Homogeneous, isotropic tetrahedron.}\\
	{\tt SHPAR1}=length/$d$ 
	of one edge. 
	Orientation: one face parallel to $y$,$z$ plane (in the TF), 
	opposite ``vertex" is in $+x$ 
	direction, and one edge is parallel to $z$ axis (in the TF).  
	Target axes ${\hat{\bf a}}_1=(1,0,0)$ [emerging from one vertex] and 
	${\hat{\bf a}}_2=(0,1,0)$ [emerging from an edge] in the TF.
	User must set {\tt NCOMP}=1.
\item {\tt TWOSPH} -- {\bf Two touching homogeneous, isotropic spheroids,
	with distinct compositions.}\\ 
	First spheroid has length {\tt SHPAR1} along symmetry axis, diameter 
	{\tt SHPAR2} perpendicular to symmetry axis.
	Second spheroid has length {\tt SHPAR3} along symmetry axis, 
	diameter {\tt SHPAR4} perpendicular to symmetry axis.  
	Contact point is on line connecting centroids.  
	Line connecting centroids is in $x$ direction.
	Symmetry axis of first spheroid is in $y$ direction.  
	Symmetry axis of second spheroid is in direction 
	${\hat{\bf y}}\cos({\tt SHPAR5})+{\hat{\bf z}}\sin({\tt SHPAR5})$,
	where ${\hat{\bf y}}$ and ${\hat{\bf z}}$ are basis vectors in TF, 
	and {\tt SHPAR5} is in degrees.  
	If {\tt SHPAR6}=0., then target axes ${\hat{\bf a}}_1=(1,0,0)$,
	${\hat{\bf a}}_2=(0,1,0)$. 
	If {\tt SHPAR6}=1., then axes ${\hat{\bf a}}_1$ and ${\hat{\bf a}}_2$ 
	are set to 
	principal axes with largest and 2nd largest moments of inertia assuming
	spheroids to be of uniform density.
	User must set {\tt NCOMP}=2 and provide dielectric function files for 
	each spheroid.
\item {\tt TWOELL} -- {\bf Two touching, homogeneous, isotropic ellipsoids,
	with distinct compositions.}\\
	{\tt SHPAR1}, {\tt SHPAR2}, {\tt SHPAR3}=x-length/$d$, y-length/$d$,
	$z$-length/$d$ of one ellipsoid.
	The two ellipsoids have identical shape, size, and orientation,
	but distinct dielectric functions.
	The line connecting ellipsoid centers is along the $x$-axis in the TF.
	Target axes ${\hat{\bf a}}_1=(1,0,0)$ 
	[along line connecting ellipsoids]
	and ${\hat{\bf a}}_2=(0,1,0)$.
	User must set {\tt NCOMP}=2 and provide dielectric function file names
	for both ellipsoids. 
	Ellipsoids are in order of increasing x:
	first dielectric function is for ellipsoid with 
	center at negative $x$, second dielectric function for 
	ellipsoid with center at positive $x$.
\item {\tt TWOAEL} -- {\bf Two touching, homogeneous, anisotropic 
	ellipsoids, with distinct compositions.}\\ 
	Geometry as for {\tt TWOELL};
	{\tt SHPAR1}, {\tt SHPAR2}, {\tt SHPAR3} have same meanings as for 
	{\tt TWOELL}.  
	Target axes ${\hat{\bf a}}_1=(1,0,0)$ and ${\hat{\bf a}}_2=(0,1,0)$
	in the TF.
	It is assumed that (for both ellipsoids) the dielectric tensor
	is diagonal in the TF.
	User must set {\tt NCOMP}=6 and provide 
	$xx$, $yy$, $zz$ components of dielectric tensor for 
	first ellipsoid, and $xx$, $yy$, $zz$ components of 
	dielectric tensor for second 
	ellipsoid (ellipsoids are in order of increasing x).
\item {\tt THRELL} -- {\bf Three touching homogeneous, isotropic ellipsoids 
	of equal size and orientation, but distinct compositions.}\\
	{\tt SHPAR1}, {\tt SHPAR2}, {\tt SHPAR3} have same meaning as for 
	{\tt TWOELL}.
	Line connecting ellipsoid centers is parallel to $x$ axis.  
	Target axis ${\hat{\bf a}}_1=(1,0,0)$ along line of ellipsoid centers, 
	${\hat{\bf a}}_2=(0,1,0)$.
	User must set {\tt NCOMP}=3 and provide (isotropic) dielectric 
	functions for first, second, and third ellipsoid.
\item {\tt THRAEL} -- {\bf Three touching homogeneous, anisotropic ellipsoids 
	with same size and 
	orientation but distinct dielectric tensors.}\\
	{\tt SHPAR1}, {\tt SHPAR2}, {\tt SHPAR3} have same meanings as for 
	{\tt THRELL}.  
	Target axis ${\hat{\bf a}}_1=(1,0,0)$ along line of ellipsoid centers, 
	${\hat{\bf a}}_2=(0,1,0)$.  
	It is assumed that dielectric tensors 
	are all diagonal in the TF.
	User must set {\tt NCOMP}=9 and provide $xx$, $yy$, $zz$ 
	elements of dielectric tensor
	for first ellipsoid, $xx$, $yy$, $zz$ elements for second ellipsoid, 
	and $xx$, $yy$, $zz$ elements for third ellipsoid 
	(ellipsoids are in order of increasing $x$).
\item {\tt BLOCKS} -- {\bf Homogeneous target constructed from 
	cubic ``blocks''.}\\  
	Number and location of blocks are specified in separate file 
	{\tt blocks.par} with following structure:\\
	\hspace*{2em}one line of comments (may be blank)\\
	\hspace*{2em}{\tt PRIN} (= 0 or 1 -- see below)\\
	\hspace*{2em}{\tt N} (= number of blocks) \\
	\hspace*{2em}{\tt B} (= width/$d$ of one block) \\
	\hspace*{2em}$x$ $y$ $z$ (= position of 1st block 
		in units of {\tt B}$d$)\\
	\hspace*{2em}$x$ $y$ $z$ (= position of 2nd block)\\
	\hspace*{2em}... \\
	\hspace*{2em}$x$ $y$ $z$ = position of {\tt N}th block\\
	If {\tt PRIN}=0, then ${\hat{\bf a}}_1=(1,0,0)$, 
	${\hat{\bf a}}_2=(0,1,0)$.
	If {\tt PRIN}=1, then ${\hat{\bf a}}_1$ and ${\hat{\bf a}}_2$ are set 
	to principal 
	axes with largest and second largest moments of inertia,
	%	\footnote{
	%	This option may not work under some compilers, in particular
	%	{\tt g77} under Linux.  This problem apparently arises
	%	in the {\tt linpack} routines used to diagonalize the
	%	moment of inertia tensor.
	%	If failure occurs, an error message will be printed.
	%	}
	assuming target to be of uniform density.
	User must set {\tt NCOMP=1}.
\item {\tt DW1996} -- {\bf 13 block target used by 
	Draine \& Weingartner (1996).} \\
	Single, isotropic material.
	Target geometry was used in study by Draine \& Weingartner (1996) 
	of radiative torques on irregular grains.
	${\hat{\bf a}}_1$ and ${\hat{\bf a}}_2$ 
	are principal axes with largest and
	second-largest moments of inertia.
	User must set {\tt NCOMP=1}.
	Target size is controlled by shape parameter {\tt SHPAR(1)} = 
	width of one block in lattice units.
\item {\tt NSPHER} -- {\bf Multisphere target consisting of the union of $N$
	spheres of single isotropic material.}\\  
	Spheres may overlap if desired.
	The relative locations and sizes of these spheres are
	defined in an external file, whose name (enclosed in quotes) 
	is passed through {\tt ddscat.par}.  The length of the file name
	should not exceed 13 characters.  
	The pertinent line in {\tt ddscat.par} should read\\
	{\tt SHPAR(1) SHPAR(2)} {\it `filename'} (quotes must be used)\\
	where {\tt SHPAR(1)} = target diameter in $x$ direction 
	(in Target Frame) in units of $d$\\
	{\tt SHPAR(2)}= 0 to have $a_1=(1,0,0)$, $a_2=(0,1,0)$ 
	in Target Frame.\\
	{\tt SHPAR(2)}= 1 to use principal axes of moment of inertia
	tensor for $a_1$ (largest $I$) and $a_2$ (intermediate $I$).\\
	{\it filename} is the name of the file specifying the locations and
	relative sizes of the spheres.\\
	The overall size of the multisphere target (in terms of numbers of
	dipoles) is determined by parameter {\tt SHPAR(1)}, which is
	the extent of the multisphere target in the $x$-direction, in
	units of the lattice spacing $d$.
	The file {\it `filename'} should have the
	following structure:\\
	\hspace*{2em}$N$ (= number of spheres)\\
	\hspace*{2em}one line of comments (may be blank)\\
	\hspace*{2em}$x_1$ $y_1$ $z_1$ $a_1$ (arb. units)\\
	\hspace*{2em}$x_2$ $y_2$ $z_2$ $a_2$ (arb. units)\\
	\hspace*{2em} ... \\
	\hspace*{2em}$x_N$ $y_N$ $z_N$ $a_N$ (arb. units)\\
	where $x_j$, $y_j$, $z_j$ are the coordinates of the center,
	and $a_j$ is the radius of sphere $j$.\\
	Note that $x_j$, $y_j$, $z_j$, $a_j$ ($j=1,...,N$) establish only
	the {\it shape} of the $N-$sphere target.  For instance,
	a target consisting of two touching spheres with the line between
	centers parallel to the $x$ axis could equally well be
	described by lines 3 and 4 being\\
	\hspace*{2em}0 0 0 0.5\\
	\hspace*{2em}1 0 0 0.5\\
	or\\
	\hspace*{2em}0 0 0 1\\
	\hspace*{2em}2 0 0 1\\
	The actual size (in physical units) is set by the value
	of $a_{\rm eff}$ specified in {\tt ddscat.par}, where, as
	always, $a_{\rm eff}\equiv (3 V/4\pi)^{1/3}$, where $V$ is the
	volume of material in the target.\\
	User must set {\tt NCOMP}=1.
\item {\tt PRISM3} -- {\bf Triangular prism of 
	homogeneous, isotropic material.}\\
	{\tt SHPAR1, SHPAR2, SHPAR3, SHPAR4} $= a/d$, $b/a$, $c/a$, $L/a$\\
	The triangular cross section has sides of width $a$, $b$, $c$.
	$L$ is the length of the prism.
	$d$ is the lattice spacing.
	The triangular cross-section 
	has interior angles $\alpha$, $\beta$, $\gamma$
	(opposite sides $a$, $b$, $c$) given by
	$\cos\alpha=(b^2+c^2-a^2)/2bc$, $\cos\beta=(a^2+c^2-b^2)/2ac$,
	$\cos\gamma=(a^2+b^2-c^2)/2ab$.
	In the Target Frame, the prism axis is in the $\hat{\bf x}$ direction,
	the normal to the rectangular face of width $a$ is (0,1,0), 
	the normal to the rectangular face of width $b$ is
	$(0,-\cos\gamma,\sin\gamma)$, and
	the normal to the rectangular face of width  $c$ is
	$(0,-\cos\gamma,-\sin\gamma)$.

\item {\tt FRMFIL} -- {\bf List of dipole locations and
	``compositions'' obtained from a file.}\\
	This option causes {{\bf DDSCAT}}\ to read the target geometry
	information from a file {\tt shape.dat} instead of automatically
	generating one of the geometries listed above.  The {\tt shape.dat}
	file is read by routine {\tt REASHP} (file {\tt reashp.f}).
	The user can customize {\tt REASHP} as needed to conform to the
	manner in which the target geometry is stored in file {\tt shape.dat}.
	However, as supplied, {\tt REASHP} expects the file {\tt shape.dat}
	to have the following structure:
	\begin{itemize}
	\item one line containing a description; the first 67 characters will
		be read and printed in various output statements.
	\item {\tt N} = number of dipoles in target
	\item $a_{1x}$ $a_{1y}$ $a_{1z}$ = x,y,z components of $\bf{a}_1$
	\item $a_{2x}$ $a_{2y}$ $a_{2z}$ = x,y,z components of $\bf{a}_2$
	\item (line containing comments)
	\item $dummy$ {\tt IXYZ(1,1) IXYZ(1,2) IXYZ(1,3) 
		ICOMP(1,1) ICOMP(1,2) ICOMP(1,3)}
	\item $dummy$ {\tt IXYZ(2,1) IXYZ(2,2) IXYZ(2,3) 
		ICOMP(2,1) ICOMP(2,2) ICOMP(2,3)}
	\item $dummy$ {\tt IXYZ(3,1) IXYZ(3,2) IXYZ(3,3)
		ICOMP(3,1) ICOMP(3,2) ICOMP(3,3)}
	\item ...
	\item $dummy$ {\tt IXYZ(J,1) IXYZ(J,2) IXYZ(J,3)
		ICOMP(J,1) ICOMP(J,2) ICOMP(J,3)}
	\item ...
	\item $dummy$ {\tt IXYZ(N,1) IXYZ(N,2) IXYZ(N,3) 
		ICOMP(N,1) ICOMP(N,2) ICOMP(N,3)}
	\end{itemize}
\end{itemize}
where the number $dummy$ is ignored, 
{\tt IXYZ(J,1-3)} = $x/d$, $y/d$, $z/d$ (TF) for $J$-th dipole,
and
{\tt ICOMP(J,1-3)} = composition identifier to specify dielectric function
appropriate for $x,y,z$ directions at location of $J$-th dipole.
For the common case of a single isotropic material,
{\tt ICOMP(J,1-3)} = 1 1 1 .

The user should be able to easily modify these routines, or write new
routines, to generate targets with other geometries.  
The user should first
examine the routine {\tt target.f} and modify it to call any new target 
generation routines desired.  
Alternatively, targets may be generated separately, and
the target description (locations of dipoles and ``composition" corresponding
to x,y,z dielectric properties at each dipole site) read in from a file by
invoking the option {\tt FRMFIL} in {\tt ddscat.f}.

It is often desirable to be able to run the target generation routines
without running the entire {{\bf DDSCAT}}\ code.  
We have therefore provided a program {\tt CALLTARGET}
which allows the user to generate targets interactively; to create this
executable just type\footnote{
	Non-Unix sites: The source code for {\tt CALLTARGET} is in the file 
	{\tt MISC.FOR}.  
	You must compile and link this to {\tt ERRMSG},
	{\tt GETSET}, {\tt LAPACKBLAS}, {\tt LAPACKSUBS}, {\tt PRINAXIS}, 
	{\tt REASHP}, {\tt TAR2EL}, {\tt TAR2SP}, {\tt TAR3EL}, 
	{\tt TARBLOCKS}, {\tt TARCEL}, 
	{\tt TARCYL}, {\tt TARELL}, {\tt TARGET}, {\tt TARHEX}, 
	{\tt TARNSP}, {\tt TARREC}, 
	{\tt TARTET},
	and {\tt WRIMSG}.  
	The source code for {\tt ERRMSG} is in {\tt SRC2.FOR}, {\tt GETSET} 
	is in {\tt SRC5.FOR}, and the rest are in {\tt SRC8.FOR} and 
	{\tt SRC9.FOR} .}\\
\indent\indent{\tt make calltarget} .\hfill\break
The program {\tt calltarget} is to be run interactively; the prompts are
self-explanatory.  
You may need to edit the code to change the device number
{\tt IDVOUT} as for {\tt DDSCAT} (see \S\ref{subsec:IDVOUT} above).

After running, {\tt calltarget} will leave behind an ASCII file 
{\tt target.out}
which is a list of the occupied lattice sites in the last target generated.
The format of {\tt target.out} is the same as the format of the {\tt shape.dat}
files read if option {\tt FRMFIL} is used (see above).
Therefore you can simply \\
\indent\indent{\tt mv target.out shape.dat} \\
and then use {\tt DDSCAT}
with the option {\tt FRMFIL}
in order to input a target shape generated by {\tt CALLTARGET}.

\section{Scattering Directions\label{sec:scattering_directions}}

{{\bf DDSCAT}}\ calculates scattering in selected directions, and elements of
the scattering matrix are reported in the output 
files {\tt w}{\it xx}{\tt r}{\it yy}{\tt k}{\it zzz}{\tt .sca} .
The scattering direction is specified through angles $\theta_s$ and $\phi_s$ 
(not to be confused with the angles $\Theta$ and $\Phi$ which specify the 
orientation of the target relative to the incident radiation!).

The scattering angle $\theta_s$ is simply the angle between the incident beam
(along direction ${\hat{\bf x}}$) and the scattered beam ($\theta_s=0$ for 
forward scattering, $\theta_s=180^\circ$ for backscattering).

The scattering angle $\phi_s$ specifies the orientation of the ``scattering 
plane'' relative to the ${\hat{\bf x}},{\hat{\bf y}}$ plane in the Lab Frame.
When $\phi_s=0$ the scattering plane is assumed to coincide with the
${\hat{\bf x}},{\hat{\bf y}}$ plane.
When $\phi_s=90^\circ$ the scattering plane is assumed to coincide with
the ${\hat{\bf x}},{\hat{\bf z}}$ plane.
Within the scattering plane the scattering directions are specified by
$0\leq\theta_s\leq180^\circ$.

Scattering directions for which the scattering properties are to
be calculated are set in the file {\tt ddscat.par} by
specifying one or more scattering planes (determined by the value of $\phi_s$)
and for each scattering plane, the number and range of $\theta_s$ values.
The only limitation is that the number of scattering directions not exceed
the parameter {\tt MXSCA} in {\tt DDSCAT.f} (in the code as distributed it
is set to {\tt MXSCA=1000}).

\section{Incident Polarization State\label{sec:incident_polarization}}

Recall that the ``Lab Frame'' is defined such that the incident radiation
is propagating along the ${\hat{\bf x}}$ axis.
{{\bf DDSCAT.5a}}\ allows the user to specify a general elliptical polarization
for the incident radiation, by specifying the (complex) polarization
vector ${\hat{\bf e}}_{01}$.
The orthonormal polarization state 
${\hat{\bf e}}_{02}={\hat{\bf x}}\times{\hat{\bf e}}_{01}^*$ is
generated automatically if {\tt ddscat.par} specifies {\tt IORTH=2}.

For incident linear polarization, one can simply set
${\hat{\bf e}}_{01}={\hat{\bf y}}$ by specifying {\tt (0,0) (1,0) (0,0)} in {\tt ddscat.par};
then ${\hat{\bf e}}_{02}={\hat{\bf z}}$
For polarization mode ${\hat{\bf e}}_{01}$ to correspond to right-handed
circular polarization, set ${\hat{\bf e}}_{01}=({\hat{\bf y}}+i{\hat{\bf z}})/\surd2$ by specifying
{\tt (0,0) (1,0) (0,1)} in {\tt ddscat.par} ({{\bf DDSCAT.5a}}\ automatically
takes care of the normalization of ${\hat{\bf e}}_{01}$); then
${\hat{\bf e}}_{02}=(i{\hat{\bf y}}+{\hat{\bf z}})/\surd2$, corresponding to left-handed
circular polarization.

\section{Averaging over Scattering: $g(1)=\langle\cos\theta_s\rangle$, 
	etc.\label{sec:averaging_scattering}}
{{\bf DDSCAT}}\ automatically carries out numerical integration of various
scattering properties.  In particular, it calculates the mean value
$g(1)=\langle\cos\theta_s\rangle$ for the scattered intensity for each
incident polarization state.
This is accomplished by evaluating the scattered intensity for
{\tt ICTHM} different values of $\theta_s$ (including $\theta_s=0$
and $\theta_s=\pi$), and taking a weighted sum.
For each value of $\theta_s$ except $0$ and $\pi$, the scattering
intensity is evaluated for {\tt IPHM} different values of the scattering
angle $\phi_s$.
The angular integration over $\theta_s$ is accomplished using Simpson's
rule (with uniform intervals in $\cos\theta_s$), 
so {\tt ICTHM} should be an {\it odd} number.
The angular integration over $\phi_s$ is accomplished by sampling
uniformly in $\phi_s$ with uniform weighting; {\tt IPHM} can be either
even or odd.

The following quantities are evaluated by this angular integration:
\begin{itemize}
\item${\bf g}=\langle\cos\theta_s\rangle{\hat{\bf x}}+
	\langle\sin\theta_s\cos\phi_s\rangle{\hat{\bf y}}+
	\langle\sin\theta_s\sin\phi_s\rangle{\hat{\bf z}}$
	(see \S\ref{sec:torque_calculation});
\item${\bf Q}_{\Gamma}$ (see \S\ref{sec:torque_calculation}).
\end{itemize}
It is important that the user recognize that accurate evaluation 
of these angular averages requires adequate sampling over scattering angles.
For small values of the size parameter $x=2\pi a_{\rm eff}/\lambda$,
the angular distribution of scattered radiation has a dipolar character
and the sampling in $\theta_s$ and $\phi_s$ does not need to be very fine,
so {\tt ICTHM} and {\tt IPHM} need not be large.
For larger values of the size parameter $x$, however, higher multipoles
in the scattered radiation field become important, and finer sampling
in $\theta_s$ and $\phi_s$ is required.
We do not have any foolproof prescription to offer, since the scattering
pattern will depend upon the target geometry and dielectric constant
in addition to overall size parameter.
However, as a very rough guide, we suggest that the user specify
values of {\tt ICTHM} and {\tt IPHM} satisfying
\begin{eqnarray}
{\tt ICTHM} &>& 5(1+x)\label{eq:ICTHM}~~~,\\
{\tt IPHM} &>& 2(1+x) ~~~.
\label{eq:IPHM}
\end{eqnarray}
The sample {\tt ddscat.par} file supplied has ${\tt ICTHM=33}$ and
${\tt IPHM}=12$; the above criteria would suggest that this would
be suitable for $x<5$.

The cpu time required for evaluation of these angular averages
is proportional to $[2+{\tt IPHM}({\tt ICTHM}-2)]$.
Since the computational time spent in evaluating these angular
integrals can be a significant part of the total, it is important
to choose values of {\tt ICTHM} and {\tt IPHM} which will provide
a suitable balance between accuracy (in this part of the overall calculation)
and cpu time.

Within one scattering plane, the scattered intensity tends to
have approximately $(1+x)$ peaks for $0\leq\theta_s\leq\pi$, so that
the above prescription for {\tt ICTHM} would have at least 5 sampling
points per maximum.
The angular distribution over $\phi_s$ is usually not as
structured as that over $\theta_s$ so we suggest that {\tt IPHM} need not
be as large as {\tt ICTHM}.
We have refrained from ``hard-wiring'' the values of {\tt ICTHM}
and {\tt IPHM} because we are not confident of the reliability of the
recommended criteria (\ref{eq:ICTHM},\ref{eq:IPHM}) -- it is up to
the user to specify appropriate values of {\tt ICTHM} and {\tt IPHM}
according to the requirements of the problem being addressed.

\section{Mueller Matrix for Scattering in Selected Directions
\label{sec:mueller_matrix}}
\subsection{Two Orthogonal Incident Polarizations ({\tt IORTH=2})}
{{\bf DDSCAT.5a}}\ internally computes the 
scattering properties of the dipole array in terms of a complex
scattering matrix $f_{ml}(\theta_s,\phi_s)$ (Draine 1988), 
where index $l=1,2$ denotes the incident polarization states,
$m=1,2$ denotes the scattered polarization state, and $\theta_s$,$\phi_s$
specify the scattering direction.
Normally {{\bf DDSCAT}}\ is used with {\tt IORTH=2} in {\tt ddscat.par}, so that
the scattering problem will be solved for both incident polarization states
($l=1$ and 2); in this subsection it will be assumed that this is the case.

Incident polarization states $l=1,2$ correspond to polarization states
${\hat{\bf e}}_{01}$, ${\hat{\bf e}}_{02}$; recall that polarization state
${\hat{\bf e}}_{01}$ is user-specified, and 
${\hat{\bf e}}_{02}=\hat{\bf x}\times{\hat{\bf e}}_{01}^*$.
Scattered polarization state $m=1$ corresponds to linear polarization of the 
scattered wave parallel to the scattering plane 
(${\hat{\bf e}}_1={\hat{\bf e}}_{\parallel s}=\hat{\theta}_s$)
and $m=2$ corresponds to linear polarization
perpendicular to the scattering plane (in the
$+\hat{\phi}_s$ direction).
The scattering matrix $f_{ml}$ was defined (Draine 1988) so that
the scattered electric field ${\bf E}_s$ is related to the incident
electric field ${\bf E}_i(0)$ at the origin (where the target is
assumed to be located) by
\begin{equation}
\left(
\begin{array}{c}
	{\bf E}_s\cdot\hat{\theta}_s\\
	{\bf E}_s\cdot\hat{\phi}_s
\end{array}
\right)
=
{\exp(i{\bf k}\cdot{\bf r})\over kr}
\left(
\begin{array}{cc}
	f_{11}&f_{12}\\
	f_{21}&f_{22}
\end{array}
\right)
\left(
\begin{array}{c}
	{\bf E}_i(0)\cdot{\hat{\bf e}}_{01}^*\\
	{\bf E}_i(0)\cdot{\hat{\bf e}}_{02}^*
\end{array}
\right) ~~~.
\label{eq:f_ml_def}
\end{equation}
The 2$\times$2 complex 
{\it amplitude scattering matrix} (with elements $S_1$, $S_2$,
$S_3$, and $S_4$) is defined so that (see Bohren \& Huffman 1983)
\begin{equation}
\left(
\begin{array}{c}
	{\bf E}_s\cdot\hat{\theta}_s\\
	-{\bf E}_s\cdot\hat{\phi}_s
\end{array}
\right)
=
{\exp(i{\bf k}\cdot{\bf r})\over -ikr}
\left(
\begin{array}{cc}
	S_2&S_3\\
	S_4&S_1
\end{array}
\right)
\left(
\begin{array}{c}
	{\bf E}_i(0)\cdot{\hat{\bf e}}_{i\parallel}\\
	{\bf E}_i(0)\cdot{\hat{\bf e}}_{i\perp}
\end{array}
\right)~~~,
\label{eq:S_ampl_def}
\end{equation}
where
${\hat{\bf e}}_{i\parallel}$, ${\hat{\bf e}}_{i\perp}$ are (real) unit vectors 
for incident polarization
parallel and perpendicular to the scattering plane (with the
customary definition of ${\hat{\bf e}}_{i\perp}={\hat{\bf e}}_{i\parallel}\times{\hat{\bf x}}$).

From (\ref{eq:f_ml_def},\ref{eq:S_ampl_def}) we may write
\begin{equation}
\left(
\begin{array}{cc}
	S_2&S_3\\
	S_4&S_1
\end{array}
\right)
\left(
\begin{array}{c}
	{\bf E}_i(0)\cdot{\hat{\bf e}}_{i\parallel}\\
	{\bf E}_i(0)\cdot{\hat{\bf e}}_{i\perp}
\end{array}
\right)
=
-i
\left(
\begin{array}{cc}
	f_{11}&f_{12}\\
	-f_{21}&-f_{22}
\end{array}
\right)
\left(
\begin{array}{c}
	{\bf E}_i(0)\cdot{\hat{\bf e}}_{01}^*\\
	{\bf E}_i(0)\cdot{\hat{\bf e}}_{02}^*
\end{array}
\right)
\label{eq:fml_rel_S_v1}
\end{equation}

Let
\begin{eqnarray}
	a&\equiv& {\hat{\bf e}}_{01}^*\cdot{\hat{\bf y}}~~~,\\
	b&\equiv& {\hat{\bf e}}_{01}^*\cdot{\hat{\bf z}}~~~,\\
	c&\equiv& {\hat{\bf e}}_{02}^*\cdot{\hat{\bf y}}~~~,\\
	d&\equiv& {\hat{\bf e}}_{02}^*\cdot{\hat{\bf z}}~~~.
\end{eqnarray}
Note that since ${\hat{\bf e}}_{01}, {\hat{\bf e}}_{02}$ could be
complex (i.e., elliptical polarization),
the quantities $a,b,c,d$ are complex.
Then
\begin{equation}
\left(
\begin{array}{c}
	{\hat{\bf e}}_{01}^*\\
	{\hat{\bf e}}_{02}^*
\end{array}
\right)
=
\left(
\begin{array}{cc}
	a&b\\
	c&d
\end{array}
\right)
\left(
\begin{array}{c}
	\hat{\bf y}\\
	\hat{\bf z}
\end{array}
\right)
\end{equation}
and eq. (\ref{eq:fml_rel_S_v1}) can be written
\begin{equation}
\left(
\begin{array}{cc}
	S_2&S_3\\
	S_4&S_1
\end{array}
\right)
\left(
\begin{array}{c}
	{\bf E}_i(0)\cdot{\hat{\bf e}}_{i\parallel}\\
	{\bf E}_i(0)\cdot{\hat{\bf e}}_{i\perp}
\end{array}
\right)
=
i
\left(
\begin{array}{cc}
	-f_{11}&-f_{12}\\
	f_{21}&f_{22}
\end{array}
\right)
\left(
\begin{array}{cc}
	a&b\\
	c&d
\end{array}
\right)
\left(
\begin{array}{c}
	{\bf E}_i(0)\cdot\hat{\bf y}\\
	{\bf E}_i(0)\cdot\hat{\bf z}
\end{array}
\right)
\label{eq:fml_rel_S_v2}
\end{equation}

The incident polarization states ${\hat{\bf e}}_{i\parallel}$ and
${\hat{\bf e}}_{i\perp}$ are related to $\hat{\bf y}$, $\hat{\bf z}$ by
\begin{equation}
\left(
\begin{array}{c}
	\hat{\bf y}\\ 
	\hat{\bf z}
\end{array}
\right)
=
\left(
\begin{array}{cc}
	\cos\phi_s&	\sin\phi_s\\
	\sin\phi_s&	-\cos\phi_s
\end{array}
\right)
\left(
\begin{array}{c}
	{\hat{\bf e}}_{i\parallel}\\
	{\hat{\bf e}}_{i\perp}
\end{array}
\right);
\label{eq:yz_vs_eis}
\end{equation}
substituting (\ref{eq:yz_vs_eis}) into (\ref{eq:fml_rel_S_v2}) we
obtain
\begin{equation}
\left(\!
\begin{array}{cc}
	S_2&S_3\\
	S_4&S_1
\end{array}
\!\right)
\left(\!
\begin{array}{c}
	{\bf E}_i(0)\cdot{\hat{\bf e}}_{i\parallel}\\
	{\bf E}_i(0)\cdot{\hat{\bf e}}_{i\perp}
\end{array}
\!\right)
=
i
\left(\!
\begin{array}{cc}
	-f_{11}&-f_{12}\\
	f_{21}&f_{22}
\end{array}
\!\right)
\left(\!
\begin{array}{cc}
	a&b\\
	c&d
\end{array}
\!\right)
\left(\!
\begin{array}{cc}
	\cos\phi_s&	\sin\phi_s\\
	\sin\phi_s&	-\cos\phi_s
\end{array}
\!\right)
\left(\!
\begin{array}{c}
	{\bf E}_i(0)\cdot\hat{\bf y}\\
	{\bf E}_i(0)\cdot\hat{\bf z}
\end{array}
\!\right)
\label{eq:fml_rel_S_v3}
\end{equation}

Eq. (\ref{eq:fml_rel_S_v3}) must be true for all ${\bf E}_i(0)$, so we
obtain an expression for the complex scattering amplitude matrix in terms
of the $f_{ml}$:
\begin{equation}
\left(
\begin{array}{cc}
	S_2&S_3\\
	S_4&S_1
\end{array}
\right)
=
i
\left(
\begin{array}{cc}
	-f_{11}&-f_{12}\\
	f_{21}&f_{22}
\end{array}
\right)
\left(
\begin{array}{cc}
	a&	b\\
	c&	d
\end{array}
\right)
\left(
\begin{array}{cc}
	\cos\phi_s&\sin\phi_s\\
	\sin\phi_s&-\cos\phi_s
\end{array}
\right)~~~.
\label{eq:fml_rel_S_v4}
\end{equation}
This provides the 4 equations used in subroutine {\tt GETMUELLER} to
compute the amplitude scattering matrix elements:
\begin{eqnarray}
S_1 &=& -i\left[
	f_{21}(b\cos\phi_s-a\sin\phi_s)
	+f_{22}(d\cos\phi_s-c\cos\phi_s)
	\right]~~~,\\
S_2 &=& -i\left[
	f_{11}(a\cos\phi_s+b\sin\phi_s)
	+f_{12}(c\cos\phi_s+d\sin\phi_s)
	\right]~~~,\\
S_3 &=& i\left[
	f_{11}(b\cos\phi_s-a\sin\phi_s)
	+f_{12}(d\cos\phi_s-c\sin\phi_s)
	\right]~~~,\\
S_4 &=& i\left[
	f_{21}(a\cos\phi_s+b\sin\phi_s)
	+f_{22}(c\cos\phi_s+d\sin\phi_s)
	\right] ~~~.
\end{eqnarray}
It is both convenient and customary to characterize both incident and
scattered radiation by 4 ``Stokes parameters'' -- 
the elements of the ``Stokes vector''.
There are different conventions in the literature; we adhere to the
definitions of the Stokes vector ($I$,$Q$,$U$,$V$) adopted in the
excellent treatise by Bohren \&
Huffman (1983), to which the reader is referred for further detail.
Here are some examples of Stokes vectors $(I,Q,U,V)$:
\begin{itemize}
\item $(1,0,0,0)I$ : unpolarized light.
\item $(1,1,0,0)I$ : 100\% linearly polarized with ${\bf E}$ parallel to the
	scattering plane;
\item $(1,-1,0,0)I$ : 100\% linearly polarized with ${\bf E}$ perpendicular
	to the scattering plane;
\item $(1,0,1,0)I$ : 100\% linearly polarized with ${\bf E}$ at +45$^\circ$
	relative to the scattering plane;
\item $(1,0,-1,0)I$ : 100\% linearly polarized with ${\bf E}$ at -45$^\circ$
	relative to the scattering plane;
\item $(1,0,0,1)I$ : 100\% right circular polarization ({\it i.e.,} negative
	helicity);
\item $(1,0,0,-1)I$ : 100\% left circular polarization ({\it i.e.,} positive
	helicity).
\end{itemize}
It is convenient to describe the scattering properties in terms of
the $4\times4$ Mueller matrix relating the Stokes parameters $(I_i,Q_i,U_i,V_i)$ and
$(I_s,Q_s,U_s,V_s)$ of
the incident and scattered radiation:
\begin{equation}
\left(
\begin{array}{c}
	I_s\\
	Q_s\\
	U_s\\
	V_s
\end{array}
\right)
=
{1\over k^2r^2}
\left(
\begin{array}{cccc}
	S_{11}&S_{12}&S_{13}&S_{14}\\
	S_{21}&S_{22}&S_{23}&S_{24}\\
	S_{31}&S_{32}&S_{33}&S_{34}\\
	S_{41}&S_{42}&S_{43}&S_{44}
\end{array}
\right)
\left(
\begin{array}{c}
	I_i\\
	Q_i\\
	U_i\\
	V_i
\end{array}
\right)~~~.
\end{equation}
This $4\times4$ Mueller matrix for scattering by a single particle
is also referred to as the
``scattering matrix'', and sometimes as the ``phase matrix''.

Once the amplitude scattering matrix elements $S_1$, $S_2$, $S_3$, and $S_4$
have been obtained, the Mueller
matrix elements can be computed (Bohren \& Huffman 1983):
\begin{eqnarray}
S_{11}&=&\left(|S_1|^2+|S_2|^2+|S_3|^2+|S_4|^2\right)/2~~~,\nonumber\\
S_{12}&=&\left(|S_2|^2-|S_1|^2+|S_4|^2-|S_3|^2\right)/2~~~,\nonumber\\
S_{13}&=&{\rm Re}\left(S_2S_3^*+S_1S_4^*\right)~~~,\nonumber\\
S_{14}&=&{\rm Im}\left(S_2S_3^*-S_1S_4^*\right)~~~,\nonumber\\
S_{21}&=&\left(|S_2|^2-|S_1|^2+|S_3|^2-|S_4|^2\right)/2~~~,\nonumber\\
S_{22}&=&\left(|S_1|^2+|S_2|^2-|S_3|^2-|S_4|^2\right)/2~~~,\nonumber\\
S_{23}&=&{\rm Re}\left(S_2S_3^*-S_1S_4^*\right)~~~,\nonumber\\
S_{24}&=&{\rm Im}\left(S_2S_3^*+S_1S_4^*\right)~~~,\nonumber\\
S_{31}&=&{\rm Re}\left(S_2S_4^*+S_1S_3^*\right)~~~,\nonumber\\
S_{32}&=&{\rm Re}\left(S_2S_4^*-S_1S_3^*\right)~~~,\nonumber\\
S_{33}&=&{\rm Re}\left(S_1S_2^*+S_3S_4^*\right)~~~,\nonumber\\
S_{34}&=&{\rm Im}\left(S_2S_1^*+S_4S_3^*\right)~~~,\nonumber\\
S_{41}&=&{\rm Im}\left(S_4S_2^*+S_1S_3^*\right)~~~,\nonumber\\
S_{42}&=&{\rm Im}\left(S_4S_2^*-S_1S_3^*\right)~~~,\nonumber\\
S_{43}&=&{\rm Im}\left(S_1S_2^*-S_3S_4^*\right)~~~,\nonumber\\
S_{44}&=&{\rm Re}\left(S_1S_2^*-S_3S_4^*\right)~~~.
\end{eqnarray}
These matrix elements are computed in {\tt DDSCAT} and passed to subroutine
{\tt WRITESCA} which handles output of scattering properties.
As delivered, {\tt WRITESCA} writes out 6 selected elements: $S_{11}$, $S_{21}$,
$S_{31}$, $S_{41}$ (these 4 elements describe the intensity and polarization 
state for scattering of unpolarized incident radiation), $S_{12}$, and $S_{13}$.
In addition, {\tt WRITESCA} writes out the linear polarization $P$ of the
scattered light for incident unpolarized light:
\begin{equation}
P = {(S_{21}^2+S_{31}^2)^{1/2}\over S_{11}} ~~~.
\end{equation}
Of course, other elements $S_{ij}$ may be of interest.
It is relatively straightforward for the user to modify subroutine
{\tt WRITESCA} to write out whatever elements of the Mueller matrix (or the
scattering amplitude matrix) are desired.
\subsection{One Incident Polarization State Only ({\tt IORTH=1})}
In some cases it may be desirable to limit the calculations to a single incident
polarization state -- for example, when each solution is very time-consuming, and
the target is known to have some symmetry so that solving for a single incident
polarization state may be sufficient for the required purpose.
In this case, set {\tt IORTH=1} in {\tt ddscat.par}.

When {\tt IORTH=1}, only $f_{11}$ and $f_{21}$ are available;
hence, {{\bf DDSCAT}}\ cannot
automatically generate the Mueller matrix elements.
In this case, the output routine {\tt WRITESCA} writes out the quantities
$|f_{11}|^2$, $|f_{21}|^2$, ${\rm Re}(f_{11}f_{21}^*)$, and
${\rm Im}(f_{11}f_{21}^*)$ for each of the scattering directions.

Note, however, that if {\tt IPHI} is greater than 1, {{\bf DDSCAT}}\ will automatically
set {\tt IORTH=2} even if {\tt ddscat.par} specified {\tt IORTH=1}: this is because
when more than one value of the target orientation angle $\Phi$ is required, there
is no additional ``cost'' to solve the scattering problem for the second incident
polarization state, since when solutions are available for 
two orthogonal states for some particular target orientation,
the solution may be obtained for another target orientation differing only in the
value of $\Phi$ by appropriate linear combinations of these solutions.
Hence we may as well solve the ``complete'' scattering problem so that we can
compute the complete Mueller matrix.

\section{Graphics and Postprocessing\label{sec:postprocessing}}

\subsection{IDL}
At present, we do not offer a comprehensive  package for {{\bf DDSCAT}}\
data postprocessing and graphical display in IDL. However, there are several
developments  worth mentioning:
First, we offer several output capabilities from within {{\bf DDSCAT}}:
ASCII (see \S\ref{subsec:ascii}), 
FORTRAN unformatted binary (see \S\ref{subsec:binary}), 
and netCDF portable binary (see \S\ref{subsec:netCDF}).
Second, we offer several skeleton IDL utilities:

\begin{itemize}
\item {\tt bhmie.pro} is our translation to IDL of the popular Bohren-Huffman
code which calculates efficiencies for spherical particles using Mie theory.

\item {\tt readbin.pro} reads the FORTRAN unformatted binary file written by
routine {\tt writebin.f}. The variables are stored in a {\tt common} block.

\item {\tt readnet.pro} reads NetCDF portable binary file and should be the
method of choice for IDL users. It offers random data access.

\item {\tt mie.pro} is an example of an interface to binary files and
the Bohren-Huffman code.  It plots a comparison of {{\bf DDSCAT}}\ results 
with scattering by equivalent radius spheres.

\end{itemize}
\noindent At present the IDL code is experimental.

\section{Miscellanea}

Additional source code, refractive index files, etc., contributed by users
will be located in the directory {\tt DDA/misc}.  
These routines and files should be
considered to be {\bf not supported} by Draine and Flatau -- 
{\it caveat receptor!}  
These routines and files should be accompanied by enough information (e.g.,
comments in source code) to explain their use.

\section{Finale}

This User Guide is somewhat inelegant, but we hope that it will
prove useful. 
The structure of the {\tt ddscat.par} file is intended to be
simple and suggestive so that, after reading the above notes once, the user
may not have to refer to them again.

The file {\tt rel\_notes} in {\tt DDA/doc} lists known bugs in 
{{\bf DDSCAT.5a}} .  
An up-to-date version will be maintained in the directory where the publicly
available source code is located (see the {\tt README} file).  
Users are encouraged to provide B. T. Draine 
({\tt draine@astro.princeton.edu}) with their
email address; bug reports, and any new releases of {{\bf DDSCAT}}, 
will be made known to those who do!

P. J. Flatau maintains the WWW page 
``SCATTERLIB - Light Scattering Codes Library"
with URL {\tt http://atol.ucsd.edu/$\sim$pflatau}.
The SCATTERLIB Internet site is a library of light scattering codes. 
Emphasis is on providing source
codes (mostly FORTRAN). However, other information related to scattering 
on spherical and
non-spherical particles is collected: an extensive list of references 
to light scattering methods, refractive index, etc. 
This URL page contains a section on the discrete dipole approximation.

\bigskip
Concrete suggestions for improving {{\bf DDSCAT}}\ (and this User Guide) are
welcomed.
If you wish to cite this User Guide, we suggest the following 
citation:

\bigskip

\noindent Draine, B.T., \& Flatau, P.J. 2000, ``User Guide for the Discrete
Dipole Approximation Code DDSCAT (Version 5a10)'',
http://xxx.lanl.gov/abs/astro-ph/0008151v2

\bigskip
Finally, the authors have one special request:
We would very much appreciate preprints and (especially!) 
reprints of any papers which make use of {{\bf DDSCAT}}!

\section{Acknowledgments}

The routine {\tt ESELF} making use of the FFT was originally written by 
Jeremy Goodman, Princeton University Observatory.  
The FFT routine {\tt FOURX} is based on a FFT routine written by Norman
Brenner (Brenner 1969). 
The routine {\tt REFICE} was written
by Steven B. Warren, based on Warren (1984). 
The routine {\tt REFWAT} was written by Eric A. Smith.  
The {\tt GPFAPACK} package was written by Clive Temperton, and
generously made available by him for use with {\tt DDSCAT}.
Subroutine {\tt CXFFT3} is based on a FFT routine
written by Clive Temperton (Temperton 1983).  
We make use of routines from the LAPACK package (Anderson {\it et al.} 1995),
the result of work by Jack Dongarra and others at the Univ. of Tennessee,
Univ. of California Berkeley, NAG Ltd., Courant Institute, Argonne National
Lab, and Rice University.
We are indebted to all of these authors for making their code available.  

We wish also to acknowledge bug reports and suggestions from {{\bf DDSCAT}}
users, including
Henrietta Lemke, Timo Nousianen, and Mike Wolff.

Development of {{\bf DDSCAT}}\ was supported in part by National Science Foundation
grants AST-8341412, AST-8612013, AST-9017082, AST-9319283,
AST-9619429 to BTD,
in part by support from the Office of Naval Research 
Young Investigator Program to PJF, and in part by DuPont Corporate Educational
Assistance to PJF.

\vfill\eject
\appendix
\section{Understanding and Modifying {\tt ddscat.par}\label{app:ddscat.par}}

In order to use DDSCAT to perform the specific calculations of interest to
you, it will be necessary to modify the {\tt ddscat.par} file.  
Here we list the sample {\tt ddscat.par} file, followed by a discussion of
how to modify this file as needed.
Note that all numerical input data in DDSCAT is read with free-format 
{\tt READ(IDEV,*)...} statements.  
Therefore you do not need to worry about the precise format in which integer 
or floating point numbers are entered on a line.
The crucial thing is that lines in {\tt ddscat.par} containing numerical 
data have the correct number of data entries, with any informational 
comments appearing {\it after} the numerical data on a given line.

{\footnotesize
\begin{verbatim}
' =================== Parameter file ===================' 
'**** PRELIMINARIES ****'
'NOTORQ'= CMTORQ*6 (DOTORQ, NOTORQ) -- either do or skip torque calculations
'PBCGST'= CMDSOL*6 (PBCGST, PETRKP) -- select solution method
'GPFAFT'= CMETHD*6 (GPFAFT, BRENNR, TMPRTN, CONVEX)
'LATTDR'= CALPHA*6 (LATTDR, LDRISO, GOBR88, DRAI88)
'NOTBIN'= CBINFLAG (ALLBIN, ORIBIN, NOTBIN)
'NOTCDF'= CNETFLAG (ALLCDF, ORICDF, NOTCDF)
'RCTNGL'= CSHAPE*6 (FRMFIL,ELLIPS,CYLNDR,RCTNGL,HEXGON,TETRAH,UNICYL,UNIELL)
8 6 4 = shape parameters PAR1, PAR2, PAR3
1         = NCOMP = number of dielectric materials
'TABLES'= CDIEL*6 (TABLES,H2OICE,H2OLIQ; if TABLES, then filenames follow...)
'diel.tab'
'**** CONJUGATE GRADIENT DEFINITIONS ****'
0       = INIT (TO BEGIN WITH |X0> = 0)
1.00e-5 = ERR = MAX ALLOWED (NORM OF |G>=AC|E>-ACA|X>)/(NORM OF AC|E>)
'**** ANGLES FOR CALCULATION OF CSCA, G'
33      = ICTHM (number of theta values for evaluation of Csca and g)
12      = IPHM (number of phi values for evaluation of Csca and g)
'**** Wavelengths (micron) ****'
6.283185 6.283185 1 'INV' = wavelengths (first,last,how many,how=LIN,INV,LOG)
'**** Effective Radii (micron) **** '
1. 1. 1  'LIN' = eff. radii (first, last, how many, how=LIN,INV,LOG)
'**** Define Incident Polarizations ****'
(0,0) (1.,0.) (0.,0.) = Polarization state e01 (k along x axis)
2 = IORTH  (=1 to do only pol. state e01; =2 to also do orth. pol. state)
1 = IWRKSC (=0 to suppress, =1 to write ".sca" file for each target orient.
'**** Prescribe Target Rotations ****'
 0.   0.  1  = BETAMI, BETAMX, NBETA (beta=rotation around a1)
 0.  90.  3  = THETMI, THETMX, NTHETA (theta=angle between a1 and k)
 0.   0.  1  = PHIMIN, PHIMAX, NPHI (phi=rotation angle of a1 around k)
'**** Specify Scattered Directions ****'
0.  0. 180. 30 = phi, thetan_min, thetan_max, dtheta (in degrees) for plane A
90. 0. 180. 30 = phi, ... for plane B
\end{verbatim}
}
\begin{tabular}{l l}
Lines	&Comments\\
1-2	&comment lines -- no need to change.\\
3	&{\tt NOTORQ} if torque calculation is not required; \\
	&{\tt DOTORQ} if torque calculation is required. \\
4	&{\tt PBCGST} is recommended; 
	other option is {\tt PETRKP} (see \S\ref{sec:choice_of_algorithm}).\\
5	&{\tt GPFAFT} recommended; other options are
	{\tt BRENNR}, {\tt TMPRTN}, {\tt CONVEX} 
	(\S\protect{\ref{sec:choice_of_fft}}).\\
6	&{\tt LATTDR} (LATTice Dispersion Relation) is recommended
	(\S\protect{\ref{sec:polarizabilities}}).\\
7	&{\tt ALLBIN} for unformatted binary dump (\S\ref{subsec:binary}; \\
	&{\tt ORIBIN} for unformatted binary dump of orientational averages only; \\
	&{\tt NOTBIN} for no unformatted binary output.\\
8	&{\tt ALLCDF} for output in netCDF format (must have netCDF option enabled; cf. \S\ref{subsec:netCDF}); \\
	&{\tt ORICDF} for orientational averages in netCDF format (must have netCDF option enabled).\\
	&{\tt NOTCDF} for no output in netCDF format; \\
9	&specify choice of target shape (see \S\ref{sec:target_generation} for
	description of options {\tt RCTNGL}, {\tt ELLIPS}, {\tt TETRAH}, ...)\\
10	&shape parameters {SHPAR1}, {SHPAR2}, {SHPAR3}, ... 
	(see \S\ref{sec:target_generation}).\\
11	&number of different dielectric constant tables (\S\ref{sec:dielectric_func}).\\
12	&name(s) of dielectric constant table(s) (one per line).\\
13	&comment line -- no need to change.\\
14	&{\tt 0} is recommended value of parameter {\tt INIT}.\\
15	&{\tt ERR} = error tolerance $h$: maximum allowed value of 
	$|A^\dagger E-A^\dagger AP|/|A^\dagger E|$ [see eq.(\ref{eq:err_tol})].\\
16	&comment line -- no need to change.\\
17	&{\tt ICTHM} -- number of $\theta_s$ values for angular averages (\S\ref{sec:averaging_scattering}).\\
18	&{\tt IPHM} -- number of $\phi_s$ values for angular averages.\\
19	&comment line -- no need to change.\\
20	&$\lambda$ -- first, last, how many, how chosen.\\
21	&comment line -- no need to change.\\
22	&$a_{\rm eff}$ -- first, last, how many, how chosen.\\
23	&comment line -- no need to change.\\
24	&specify x,y,z components of (complex) incident polarization ${\hat{\bf e}}_{01}$ (\S\ref{sec:incident_polarization})\\
25	&{\tt IORTH} = 1 to do one polarization state only;\\
	&2 to do second (orthogonal) incident polarization as well.\\
26	&{\tt IWRKSC} = 0 to suppress writing of ``.sca" files;\\
	&2 to enable writing of ``.sca" files.\\
27	&comment line -- no need to change.\\
28	&$\beta$ (see \S\ref{sec:target_orientation}) -- first, last, how many .\\
29	&$\Theta$ -- first, last, how many.\\
30	&$\Phi$ -- first, last, how many.\\
31	&comment line -- no need to change.\\
32	&$\phi_s$ for first scattering plane, $\theta_{s,min}$, $\theta_{s,max}$, how many $\theta_s$ values;\\
33,...	&$\phi_s$ for 2nd,... scattering plane, ...
\end{tabular}

\newpage
\section{{\tt w{\it xx}r{\it yy}ori.avg} Files\label{app:w00r00ori.avg}}

The file {\tt w00r00ori.avg} contains the results for the first wavelength
({\tt w00}) and first target radius ({\tt r00})
averaged over orientations ({\tt ori.avg}).
The {\tt w00r00ori.avg} file generated by the sample calculation should look 
like the following:
{\footnotesize
\begin{verbatim}
 DDSCAT --- DDSCAT.5a.10 [00.11.03]
 TARGET --- Rectangular prism; NX,NY,NZ=   8   6   4                          
 GPFAFT --- method of solution 
 LATTDR --- prescription for polarizabilies
 RCTNGL --- shape 
 NAT0 =     192 = number of dipoles
  AEFF=   1.00000 = effective radius (physical units)
  WAVE=   6.28319 = wavelength (physical units)
   K*A=   1.00000 = 2*pi*aeff/lambda
 n= ( 1.3300 ,  0.0100),  epsilon= (  1.7688 ,  0.0266)  for material 1
   TOL= 1.000E-05 = error tolerance for CCG method
 ICTHM= 33 = theta values used in comp. of Qsca,g
 IPHIM= 12 = phi values used in comp. of Qsca,g
 ( 1.00000  0.00000  0.00000) = target axis A1 in Target Frame
 ( 0.00000  1.00000  0.00000) = target axis A2 in Target Frame
 ( 0.27942  0.00000  0.00000) = k vector (latt. units) in Lab Frame
 ( 0.00000, 0.00000)( 1.00000, 0.00000)( 0.00000, 0.00000)=inc.pol.vec. 1 in LF
 ( 0.00000, 0.00000)( 0.00000, 0.00000)( 1.00000, 0.00000)=inc.pol.vec. 2 in LF
   0.000   0.000 = beta_min, beta_max ;  NBETA = 1
   0.000  90.000 = theta_min, theta_max; NTHETA= 3
   0.000   0.000 = phi_min, phi_max   ;   NPHI = 1
 Results averaged over   3 target orientations
                   and   2 incident polarizations
          Qext      Qabs       Qsca     g=<cos>    Qbk       Qpha
 JO=1:  1.330E-01 3.269E-02 1.003E-01  2.345E-01 5.552E-03 4.836E-01
 JO=2:  9.063E-02 2.397E-02 6.666E-02  2.673E-01 3.430E-03 4.159E-01
 mean:  1.118E-01 2.833E-02 8.347E-02  2.476E-01 4.491E-03 4.497E-01
 Qpol=  4.234E-02                                   dQpha= 6.768E-02
        Qsca*g(1)  Qsca*g(2)  Qsca*g(3)
 JO=1:  2.352E-02  1.126E-03 -7.190E-10
 JO=2:  1.781E-02  2.768E-03  3.244E-11
 mean:  2.066E-02  1.947E-03 -3.433E-10
            ** Mueller matrix elements for selected scattering directions **
theta   phi   S_11       S_21       S_31       S_41       S_12       S_13      Pol.
  0.0   0.0 5.167E-02  7.916E-03  9.582E-11  1.267E-11  7.916E-03  9.239E-11  0.15320
 30.0   0.0 4.046E-02  4.518E-04  2.540E-11  4.619E-11  4.518E-04  2.339E-11  0.01117
 60.0   0.0 2.169E-02 -1.023E-02  8.752E-11  5.628E-11 -1.023E-02  6.967E-11  0.47184
 90.0   0.0 1.193E-02 -1.192E-02  2.068E-11  6.778E-11 -1.192E-02 -1.481E-11  0.99888
120.0   0.0 1.170E-02 -5.702E-03 -4.564E-11  3.005E-11 -5.702E-03  5.392E-11  0.48743
150.0   0.0 1.403E-02  9.758E-04 -2.008E-11 -1.155E-11  9.758E-04  1.665E-11  0.06956
180.0   0.0 1.411E-02  3.333E-03 -1.390E-11  1.454E-11  3.333E-03  1.643E-11  0.23621
  0.0  90.0 5.167E-02 -7.916E-03 -7.574E-10  4.831E-11 -7.916E-03 -7.428E-10  0.15320
 30.0  90.0 4.487E-02 -1.280E-02 -5.036E-04  8.979E-05 -1.281E-02 -3.843E-04  0.28550
 60.0  90.0 3.040E-02 -2.062E-02 -7.990E-04  1.643E-04 -2.063E-02 -3.773E-04  0.67857
 90.0  90.0 1.991E-02 -1.989E-02 -7.361E-04  1.827E-04 -1.991E-02 -3.807E-05  0.99926
120.0  90.0 1.626E-02 -1.175E-02 -4.415E-04  1.359E-04 -1.176E-02  1.634E-04  0.72285
150.0  90.0 1.478E-02 -5.293E-03 -1.716E-04  6.554E-05 -5.295E-03  1.131E-04  0.35833
180.0  90.0 1.411E-02 -3.333E-03  3.067E-10 -2.982E-11 -3.333E-03 -3.009E-10  0.23621
\end{verbatim}
}
\section{{\tt w{\it xx}r{\it yy}k{\it zzz}.sca} Files
	\label{app:w00r00k000.sca}}

The {\tt w00r00k000.sca} file contains the results for the first
wavelength ({\tt w00}), first target radius ({\tt r00}),
and first orientation ({\tt k000}).
The {\tt w00r00k000.sca} file created by the sample calculation should look
like the following:
{\footnotesize
\begin{verbatim}
 DDSCAT --- DDSCAT.5a.10 [00.11.03]
 TARGET --- Rectangular prism; NX,NY,NZ=   8   6   4                          
 GPFAFT --- method of solution 
 LATTDR --- prescription for polarizabilies
 RCTNGL --- shape 
 NAT0 =     192 = number of dipoles
  AEFF=   1.00000 = effective radius (physical units)
  WAVE=   6.28319 = wavelength (physical units)
   K*A=   1.00000 = 2*pi*aeff/lambda
 n= ( 1.3300 ,  0.0100),  epsilon= (  1.7688 ,  0.0266)  for material 1
   TOL= 1.000E-05 = error tolerance for CCG method
 ICTHM= 33 = theta values used in comp. of Qsca,g
 IPHIM= 12 = phi values used in comp. of Qsca,g
 ( 1.00000  0.00000  0.00000) = target axis A1 in Target Frame
 ( 0.00000  1.00000  0.00000) = target axis A2 in Target Frame
 ( 0.27942  0.00000  0.00000) = k vector (latt. units) in TF
 ( 0.00000, 0.00000)( 1.00000, 0.00000)( 0.00000, 0.00000)=inc.pol.vec. 1 in TF
 ( 0.00000, 0.00000)( 0.00000, 0.00000)( 1.00000, 0.00000)=inc.pol.vec. 2 in TF
 BETA =  0.000 = rotation of target around A1
 THETA=  0.000 = angle between A1 and k
  PHI =  0.000 = rotation of A1 around k
          Qext      Qabs       Qsca     g=<cos>    Qbk       Qpha
 JO=1:  1.110E-01 3.028E-02 8.077E-02  3.504E-01 1.858E-03 4.668E-01
 JO=2:  8.651E-02 2.441E-02 6.209E-02  3.650E-01 1.362E-03 4.197E-01
 mean:  9.878E-02 2.735E-02 7.143E-02  3.567E-01 1.610E-03 4.432E-01
 Qpol=  2.454E-02                                   dQpha= 4.711E-02
        Qsca*g(1)  Qsca*g(2)  Qsca*g(3)
 JO=1:  2.830E-02  9.829E-10 -2.070E-09
 JO=2:  2.266E-02  1.827E-09  9.086E-11
 mean:  2.548E-02  1.405E-09 -9.896E-10
            ** Mueller matrix elements for selected scattering directions **
theta   phi   S_11       S_21       S_31       S_41       S_12       S_13      Pol.
  0.0   0.0 4.987E-02  5.372E-03  8.485E-11 -1.807E-11  5.372E-03  8.482E-11  0.10772
 30.0   0.0 4.040E-02 -9.275E-04 -4.151E-11  1.973E-11 -9.275E-04 -4.057E-11  0.02296
 60.0   0.0 2.191E-02 -1.060E-02  1.292E-10  1.107E-11 -1.060E-02  1.205E-10  0.48383
 90.0   0.0 1.058E-02 -1.056E-02 -2.798E-11  7.613E-11 -1.056E-02 -1.444E-10  0.99794
120.0   0.0 7.506E-03 -3.995E-03  3.887E-11 -3.829E-11 -3.995E-03 -4.776E-12  0.53222
150.0   0.0 5.921E-03 -9.862E-06  1.709E-11 -8.195E-13 -9.862E-06 -1.717E-11  0.00167
180.0   0.0 5.058E-03  7.791E-04  6.177E-12 -8.607E-12  7.791E-04 -4.347E-12  0.15403
  0.0  90.0 4.987E-02 -5.372E-03 -4.752E-10  4.820E-11 -5.372E-03 -5.005E-10  0.10772
 30.0  90.0 4.285E-02 -1.032E-02 -5.283E-10 -1.079E-11 -1.032E-02 -5.098E-10  0.24082
 60.0  90.0 2.736E-02 -1.773E-02 -2.169E-10  7.855E-11 -1.773E-02 -2.703E-10  0.64798
 90.0  90.0 1.540E-02 -1.539E-02 -2.506E-11  2.845E-11 -1.539E-02 -1.588E-10  0.99942
120.0  90.0 9.786E-03 -6.745E-03  1.827E-11  3.636E-12 -6.745E-03 -7.262E-11  0.68918
150.0  90.0 6.389E-03 -1.820E-03  4.853E-11  2.225E-11 -1.820E-03 -5.195E-11  0.28491
180.0  90.0 5.058E-03 -7.791E-04  5.552E-11  2.060E-12 -7.791E-04 -5.362E-11  0.15403
\end{verbatim}
}
\end{document}